\documentclass[lettersize,journal]{IEEEtran}
\usepackage{amsmath,amsfonts}
\usepackage{amsthm}
\usepackage{array}
\usepackage[caption=false,font=normalsize,labelfont=sf,textfont=sf]{subfig}
\usepackage{textcomp}
\usepackage{stfloats}
\usepackage{url}
\usepackage{verbatim}
\usepackage{graphicx}
\usepackage{cite}
\usepackage{diagbox}

\usepackage[colorlinks,linkcolor=blue, anchorcolor=blue, citecolor=blue]{hyperref}
\usepackage{multirow}
\usepackage{amsmath,amssymb,amsfonts}
\usepackage{textcomp}
\usepackage[noend]{algpseudocode}
\usepackage{algorithmicx,algorithm}
\usepackage{cases}
\usepackage{bm}
\usepackage{color,xcolor}
\usepackage[OT2,OT1]{fontenc}

\allowdisplaybreaks[2]

\newtheorem{lemma}{Lemma}
\newtheorem{theorem}{Theorem}

\newcommand{\MYlefttab}{\!\!\!\!\!\!\!\!\!\!\!\!\!\!\!\!\!\!\!\!}

\begin{document}

\title{PeF: \underline{P}oisson's \underline{E}quation Based Large-Scale  Fixed-Outline \underline{F}loorplanning}

\author{ Ximeng Li, Keyu Peng, Fuxing Huang and Wenxing Zhu
	\thanks{
		The authors are with the Center for Discrete Mathematics and Theoretical Computer Science, Fuzhou University, Fuzhou 350108, China (Corresponding author: Wenxing Zhu, e-mail: wxzhu@fzu.edu.cn).
	}
}



\maketitle

\begin{abstract}
	Floorplanning is the first stage of VLSI physical design. An effective floorplanning engine definitely has positive impact on chip design speed, quality and performance. In this paper, we present a novel mathematical model to characterize non-overlapping of modules, and propose a flat fixed-outline floorplanning algorithm based on the VLSI global placement approach using Poisson's equation. The algorithm consists of global floorplanning and legalization phases. In global floorplanning, we redefine the potential energy of each module based on the novel mathematical model for characterizing non-overlapping of modules and an analytical solution of  Poisson's equation. In this scheme, the widths of soft modules appear as variables in the energy function and can be optimized.  Moreover, we design a fast approximate computation scheme for partial derivatives of the potential energy. In legalization, based on the defined horizontal and vertical constraint graphs, we eliminate overlaps between modules remained after global floorplanning, by  modifying relative positions of modules.
	Experiments on the MCNC, GSRC, HB+ and ami49\_x benchmarks show that, our algorithm improves the average wirelength by at least 2\% and 5\%
	on small and large scale benchmarks with certain whitespace, respectively, compared to state-of-the-art floorplanners.
\end{abstract}


\section{Introduction}
\label{lab:yy}

\IEEEPARstart{F}{loorplanning}  is the first stage of VLSI physical design flow.
With  widespread usage of IP cores and huge number of cells  integrated in modern chips, floorplanning needs to handle hundreds or even thousands of circuit modules within a fixed-outline. Especially, some more complex issues need to be considered during floorplanning, such as timing, thermal, power and voltage-island  \cite{2014F-FM,2018LJM}. However, it is well known that the fundamental floorplanning problem is  NP-hard. Hence
designing a high performance algorithm for the fundamental large-scale floorplanning is a  challenge,
which will have an impact on  the performance and  yield of the final  ICs.
In this paper, we focus on designing a flat analytical engine for the fixed-outline floorplanning.

Fixed-outline floorplanning can be described as placing
a set of rectangular modules into a fixed rectangular area without overlap,
while  minimizing the total wirelength among the modules.
Floorplanning usually includes two types of circuit modules,
hard modules and soft modules.
The width and height of a soft module can be changed
within a certain range meanwhile maintaining a certain area.
So far, works for fixed-outline floorplanning can be divided into  three categories: heuristic methods for small-scale problem,
multilevel heuristic methods for large-scale problem,
and analytical approaches.

Heuristic methods for small-scale fixed-outline floorplanning problem first adopt a floorplan representation,
then use heuristic algorithms to search for an optimal solution in the representation space,
and finally decode the  optimal representation to the corresponding floorplan.
Floorplan representation method also determines the complexity of transforming
a floorplan into its representation,
and the complexity of decoding a representation into its corresponding floorplan.
Floorplan representation methods can be summarized in Table \ref{tab:show}.

\begin{table}[h]\centering
	\caption{Summary of floorplan representations}
	\label{tab:show}
	\resizebox{\hsize}{!}{
		\begin{tabular}{|l|l|l|c|}
			\hline
			Representation                     & Solution Space          & Packing Time & Flexibility \\ \hline
			Normalized Polish Expression       & $O(n!2^{2.6n}/n^{1.5})$ & $O(n)$       & slicing     \\ \hline
			Corner Block List\cite{2000CBL}    & $O(n!2^{3n})$           & $O(n)$       & Mosaic      \\ \hline
			Twin Binary Sequence\cite{2003TBS} & $O(n!2^{3n}/n^{1.5})$   & $O(n)$       & Mosaic      \\ \hline
			O-tree\cite{1999OTree}             & $O(n!2^{2n}/n^{1.5})$   & $O(n)$       & compacted   \\ \hline
			B*-tree\cite{2000BTree}            & $O(n!2^{2n}/n^{1.5})$   & $O(n)$       & compacted   \\ \hline
			Corner Sequence\cite{2003CS}       & $\le(n!)^2$             & $O(n)$       & compacted   \\ \hline
			Sequence Pair\cite{1995SP}         & $(n!)^2$                & $O(n^2)$     & general     \\ \hline
			TCG-S\cite{2002TCGS}               & $(n!)^2$                & $O(n\log n)$ & general     \\ \hline
			ACG\cite{2004ACG}                  & $O((n!)^2)$             & $O(n^2)$     & general     \\ \hline
		\end{tabular}
	}
\end{table}

In Table  \ref{tab:show},
O-tree \cite{1999OTree} and B*-tree \cite{2000BTree} are two commonly used floorplan representations, followed by the most preferred simulated annealing algorithm.
Parquet \cite{2003Fixed} used sequence pair as the representation, on which a floorplan is also optimized using simulated annealing.
Since the running time of this kind of algorithms grows very fast with the problem scale, they cannot be applied to large-scale floorplanning directly.
A methodology to cope with this issue is the multilevel framework with the heuristic search methods.

The core idea of multilevel heuristic search methods is  reducing the size of solution space multilevelly by  clustering or partitioning, and thus can handle large-scale problem.
Clustering-based floorplanning methods, e.g., MB*-tree \cite{2003Multilevel}, merge modules from the bottom to up,
until the number of modules is small enough, and then solve the floorplanning problem.
The solution is refined  in each level of de-clustering, leading finally to a solution of the original  problem.
A disadvantage of  clustering-based floorplanning is that, modules cannot be clustered from a global perspective in the early clustering stage, which limits the solution quality.
Instead, partitioning-based methods
decompose the original large-scale floorplanning problem into
several small-scale sub-problems, until they are small enough and solved directly. Finally, solutions of sub-problems are merged and optimized level by level,  leading to a solution of the original problem.
Capo \cite{2004capo10.2}, DeFer \cite{Yan2010DeFer},
IMF \cite{2008A} and QinFer \cite{2021QinFer} are
state-of-the-art partitioning-based floorplanning methods.	Most of  them use hMetis \cite{hMetis} for the partitioning.
Compared to clustering-based floorplanning methods, partitioning-based ones can consider  wirelength information in the early floorplanning stage. However, they cannot accurately calculate the wirelength in the partitioning stage.

Since floorplanning is similar to the VLSI placement problem in some degree,
a number of analytical floorplanning methods have been proposed based on the frameworks of VLSI placement approaches. They
include  Analytical \cite{2006AFD}, AR \cite{2008Large},
UFO \cite{2011UFO} and F-FM \cite{2014F-FM}, etc.
Generally, these methods analogously have the
global floorplanning stage which determines ``rough" positions of modules, and a legalization stage which eliminates the overlaps between modules and determines the exact positions of all modules and the widths of  soft modules.
Furthermore, combining with multilevel framework, these  analytical methods can solve large-scale floorplanning problems well.
Table \ref{tab:A_M} summarizes the global floorplanning
and legalization stages of the above methods, respectively.
\begin{table}[h]\centering
	\caption{Summary of  analytical floorplanning methods}
	\label{tab:A_M}
	\resizebox{\hsize}{!}{
		\begin{tabular}{|c|c|c|}
			\hline
			method                    & global floorplanning                      & legalization                            \\ \hline
			Analytical\cite{2006AFD}  &
			\begin{tabular}[c]{@{}c@{}}
				$\operatorname{bell-shaped}$
				smoothing \\and Density control
			\end{tabular} &
			function $pl2sp()$ in Parquet-4\cite{2003Fixed}                                                                 \\
			\hline
			F-FM\cite{2014F-FM}       &
			\begin{tabular}[c]{@{}c@{}}placement model \\
				based on NTUplace\cite{2008NTUplace3}\end{tabular} &
			\begin{tabular}[c]{@{}c@{}}
				gradual partition based on \\position and area $+$ ST-Tree \cite{1986ST-tree}
			\end{tabular}                                                                                       \\ \hline
			Ref.\cite{2018LJM}        &
			\begin{tabular}[c]{@{}c@{}}placement model \\
				based on NTUplace\cite{2008NTUplace3}
			\end{tabular}
			                          & SAINT\cite{2016SAINT}                                                               \\ \hline
			AR\cite{2008Large}        & $\operatorname{Attractor-Repeller}$ model & SOCP  based on relative position matrix \\ \hline
			UFO\cite{2011UFO}         & Push-Pull model                           & SOCP based on  constraint graph         \\ \hline
		\end{tabular}
	}
\end{table}

Floorplanners Analytical \cite{2006AFD}, F-FM \cite{2014F-FM}
and Ref. \cite{2018LJM} use density-control in the global floorplanning stage.
In Analytical \cite{2006AFD}, cell density is calculated according to the
current positions and widths of modules, in which  soft modules are treated as hard ones. Then the density function is smoothed using the bell-shaped function.
During legalization, overlaps between modules are removed using
the  function $pl2sp()$ in Parquet-4\cite{2003Fixed} to get a legal floorplan.
Similarly, F-FM \cite{2014F-FM} and Ref. \cite{2018LJM}  treat all soft  modules as hard blocks,
and use global placement  for global floorplanning.
In legalization,  F-FM  \cite{2014F-FM} uses the ST-tree  \cite{1986ST-tree}
with the gradual partition based on positions and areas of modules,	
and then adopts the simulated annealing algorithm to optimize
the positions and widths of  modules.
As for the floorplanner in Ref.\cite{2018LJM}, SAINT \cite{2016SAINT} is used for legalization.
The main idea is
using polygon to approximate the curve of area of every soft module,
and building an ILP model to achieve legalization of modules.



Earlier floorplanners AR \cite{2008Large} and UFO \cite{2011UFO} directly regard
all soft and hard modules as squares, and further abstract  them as circles.
Then, Attractor-Repeller and Push-Pull models are proposed for global floorplanning respectively.
Both of them  use  second-order cone programming (SOCP)
to obtain legal solutions in the legalization stage.
The difference is that, 
AR uses relative position matrix, while UFO uses  constraint graph.
It must be remarked that solving SOCP is rather time-consuming
when the problem size is moderate or large.


Note that floorplanning problem contains hard and soft modules.
However,  state-of-the-art analytical floorplanners Analytical\cite{2006AFD},
F-FM \cite{2014F-FM}  and Ref. \cite{2018LJM}
take soft modules as hard ones in  global floorplanning. Moreover, they must be incorporated in the multilevel framework  to handle large-scale floorplanning problem. Apparently, these will affect the solution quality.





In this paper, we focus on the fundamental fixed-outline floorplanning problem.
This is because algorithms for floorplanning with complex issues
are extended from those for the fundamental floorplanning.
Notice that Poisson's equation based VLSI global placement methods \cite{2015ePlace, Pplace}
do not use multilevel framework.
They are flat while achieve state-of-the-art results.
In this paper, we will propose a high performance  floorplanning engine by extending this type of methods to large-scale fixed-outline floorplanning with hard and soft modules.

To achieve this goal, we must not only consider  optimizing
the shapes of soft modules in  global floorplanning,
but also design a high performance legalization algorithm correspondingly.
Contributions of this paper are summarized as follows:

\begin{itemize}
	\item We propose a novel mathematical model to characterize non-overlapping of modules.
	      Then under the concept of electrostatic system \cite{2015ePlace},
	      we redefine the potential energy of modules based on
	      the analytical solution of Poisson's equation in \cite{Pplace}.
	      This allows the widths of soft modules to appear  as  variables in the
	      established fixed-outline global floorplanning model,
	      and the derivative with respect to the widths of soft modules can be easily obtained.

	\item We give a fast approximate computation scheme for the partial derivative of the potential energy,
	      and design a fixed-outline global floorplanning algorithm based on the global placement method using Poisson's equation.

	\item By defining horizontal and vertical constraint graphs, we present a fast legalization algorithm to remove overlaps between modules after global floorplanning. The algorithm  modifies relative positions of modules and compresses the floorplan to obtain finally a legal solution.

	\item Within acceptable running time, our floorplanning algorithm obtains
	      better results on  large-scale benchmarks. 
	      The average wirelength is at least 5\% less than state-of-the-art multilevel heuristic search algorithms,
	      and is at least 8\% less than state-of-the-art analytical floorplanning algorithms. On small-scale benchmarks, 
	      it has similar improvements.
\end{itemize}


The remainder of this paper is organized as follows. Section \ref{lab:ybzs} is the preliminaries.
Section \ref{lab:mxjlyqj} proposes a novel mathematical model to characterize non-overlapping of modules, gives our global floorplanning model based on  Poisson's equation,
and provides a fast computation scheme for the partial derivatives.
In Section \ref{lab:sf}, we detail  our fixed-outline global floorplanning algorithm  and
legalization algorithm. Experimental results  and comparisons are presented in
Section \ref{lab:syjg}, and conclusion is summarized in Section \ref{conclu}.

\section{Preliminaries}\label{lab:ybzs}

This section presents the problem of fixed-outline floorplanning, introduces the wirelength model and  principle of the Poisson's equation method for handling modules overlap in VLSI global placement.

\subsection{ Fixed-outline Floorplanning Problem}\label{lab:wtmx}

The floorplanning area is a rectangle with four edges parallel to the coordinate axes, with the lower left corner at the origin, and upper right corner at $(W, H)$. Let $V=V_s\cup V_h$ be the set of rectangular modules, where $V_s$ and $V_h$ represent the sets of soft and hard modules, respectively.
For each module $v_i\in V$, its width, height, area and the coordinate of the center point are $w_i$, $h_i$, $A_i$ and $(x_i, y_i)$, respectively.
For each module $v_i\in V_s$, the aspect ratio, calculated as the ratio of height to width, is between $AR_i^l$ and $AR_i^u$.
Moreover, let $E$ be the set of nets $e_i$, $i=1, 2, \ldots, m$, for which $WL(e_i)$ is the half-perimeter wirelength.
Let $WL(E)=\sum_{e_i\in E}WL(e_i)$. Then the fixed-outline floorplanning problem can be described as
\begin{equation}\label{mod:1}
	\begin{aligned}
		\min \quad & WL(E)                                                 \\
		s.t. \quad & \mbox{overlap}=0                                      \\
		           & \mbox{aspect\ ratio\ is\ suitable}                    \\
		           & \mbox{all\ modules\ are\ within\ the\ fixed-outline}.
	\end{aligned}
\end{equation}

In this paper, the objective of model \eqref{mod:1} adopts the LSE (Log-Sum-Exp) wirelength function. 
Moreover, the first constraint in  \eqref{mod:1} ensures that all modules do not overlap.
The second constraint means that the aspect ratio of every soft module meets the upper and lower bounds. The third constraint indicates that all modules are located within the fixed-outline.

\subsection{VLSI Global Placement Based on Poisson's Equation}\label{lab:bsfc}

ePlace \cite{2015ePlace} is a representative VLSI global placement method. In ePlace, each module $v_i$ is regarded as a positively charged particle with electric quantity $q_i$, which is proportional to its area $A_i$. Thus the placement problem is modeled as a two-dimensional electrostatic
system,
in which the electric potential $\psi(x,y)$
and the electric field $\bm{\xi}(x,y)$ satisfy
that $\bm{\xi}(x,y)=(\xi_x,\xi_y)=-\nabla \psi(x,y)$.
ePlace \cite{2015ePlace} characterizes $\psi(x,y)$ as the solution of the Poisson's equation 	
\begin{subnumcases}{\label{eq:poisson}}
	\nabla^2\psi(x,y)=-\rho(x,y),    &$(x,y)\in \bm{R}  $,   \label{eq:poisson:1} \\
	\hat{\bm{n}}\nabla\psi(x,y)=0, & $(x,y)\in \partial\bm{R}$, \label{eq:poisson:2} \\	\iint_{\bm{R}}\rho(x,y)\mathrm{d}x\mathrm{d}y=\iint_{\bm{R}}\psi(x,y)\mathrm{d}x\mathrm{d}y=0,
	\MYlefttab\MYlefttab\MYlefttab& \label{eq:poisson:3}
\end{subnumcases}
where $\bm{R}$ is the rectangular placement area.
 Eq. \eqref{eq:poisson:1} is the Poisson's equation, $\rho(x,y)$ is the charge density function on $\bm{R}$. Eq. \eqref{eq:poisson:2} is the Neumann boundary condition,
where $\partial\bm{R}$ and $\hat{\bm{n}}$ represent the
boundary of area $\bm{R}$ and the outer normal vector of the boundary, respectively. Eq. \eqref{eq:poisson:3} is the compatibility condition to ensure  that the equation has a unique solution.

The  electric forces $\bm{F_i}=q_i\bm{\xi(x_i, y_i)}$, 
$i=1, 2,\ldots, n$, guide the modules movement and henceforth reduce modules overlaps. Define the system potential energy 
$N(V)=\sum_{v_i\in V}N(v_i)$, where $N(v_i)=q_i\psi(v_i)$ is the potential energy of module $v_i$.  Then by ePlace \cite{2015ePlace}, a necessary condition for  $overlap=0$ is that the system potential energy $N(V)$ is the smallest. This allows to use the regularizing term $N(V)$ to transform the VLSI global placement problem into 
\begin{equation}\label{mod:2}
	\begin{aligned}
		\min\quad & LSE(E) + \lambda N(V)                               \\
		s.t.\quad & \mbox{all\ modules\ are\ in\ the\ placement\ area}.
	\end{aligned}
\end{equation}

Ref. \cite{Pplace} defines the density function 
\begin{equation}\label{eq:poisson:rho_l}
	\begin{aligned}
		\rho(x,y)   & =  \sum_{v_i\in V}\rho_i(x,y), \quad (x,y)\in [0,W]\times [0,H], \\
		\rho_i(x,y) & =  \begin{cases}
			1, & if\ (x,y)\in R_i; \\
			0, & else,
		\end{cases}
	\end{aligned}
\end{equation}
where $ R_i =  \left[x_i-\frac{w_i}{2},x_i+\frac{w_i}{2}\right]
	\times\left[y_i-\frac{h_i}{2},y_i+\frac{h_i}{2}\right],
	h_i=\frac{A_i}{w_i}$.  Then by redefining
$$\rho(x, y)\triangleq \rho(x, y)- \frac 1{WH} \iint_{\bm{R}}\rho(x,y)\mathrm{d}x\mathrm{d}y,$$
Ref. \cite{Pplace} obtains an analytical solution of Eq. \eqref{eq:poisson}. The truncated version is
\begin{equation}\label{eq:poisson:solver}
	\begin{aligned}
		\psi(x,y)=\sum_{u=0}^{K}\sum_{p=0}^{K}a_{up}
		\cos\left(\frac{u\pi}{W}x\right)
		\cos\left(\frac{p\pi}{H}y\right),
	\end{aligned}
\end{equation}
where $K$ is the truncation factor that controls the precision.

For fast global placement, assume that the placement area is divided in to $K\times K$ bins, and the density of each grid is 
		\begin{equation}\label{eq:p:b}
			\begin{aligned}
				 & \hat{\rho}(i, j)  = \sum_{v_k\in V}\frac{Area(bin_{i,j}\cap R_k)}{Area(bin_{i,j})}
				,
				~i,j={0,1,\ldots, K-1}.
			\end{aligned}
		\end{equation}

By a similar redefinition, Ref. \cite{Pplace} obtains an analytical solution of Eq. \eqref{eq:poisson}, for which
the  truncated solution value  at $(i+\frac{1}{2}, j+\frac{1}{2})$ is
		\begin{equation}\label{eq:aup:new}
			\hat{\psi}(i,j)=\!\!\sum_{u,p=0}^{K-1}
			\!\!a_{up}
			\cos\left(\frac{u(i+\frac{1}{2})\pi}{K}\right)
			\cos\left(\frac{p(j+\frac{1}{2})\pi}{K}\right).
		\end{equation}
The coefficient calculations for Eqs. \eqref{eq:poisson:solver} and \eqref{eq:aup:new} can be found in \cite{Pplace}.



\section{Global Floorplanning Model and Gradient Calculation}\label{lab:mxjlyqj}

In this paper, the fixed-outline floorplanning is decomposed into two stages: global floorplanning and legalization. In global floorplanning, partial overlaps between modules are allowed and a ``rough" floorplanning is built. In  legalization, the overlaps are eliminated and the final legal floorplan is obtained. To be able to optimize  the shapes of soft modules in  global floorplanning, we propose in this section a novel mathematical model to characterize non-overlapping of modules, and then modify the VLSI global placement model \eqref{mod:2} for fixed-outline global floorplanning. Based on the global floorplanning model, we present a fast approximate scheme for calculating partial derivatives of  potential energy.


\subsection{Global Floorplanning Model}\label{lab:mxjl}


First, we present a sufficient and necessary condition for non-overlapping of modules. According  to Eq. \eqref{eq:poisson:rho_l}, we can prove the following results. The details  will be clarified in the \nameref{appndx}.

\begin{lemma}\label{lem:1}
	Any two different modules $v_i$ and $v_j\in V$ are non-overlapping
	if and only if $\iint_{R_i}\rho_j(u,v)\mathrm{d}u \mathrm{d}v=0$.
\end{lemma}

Let $Area(R_i)$ be the area of module $v_i$, we have
\begin{lemma}\label{lem:2}
	$\iint_{R_i}\rho_i(u,v)\mathrm{d}u \mathrm{d}v=Area(R_i)$  for all modules $v_i\in V$.
\end{lemma}
\begin{theorem}\label{thm:1}
	Any module $v_i\in V$ is non-overlapping with other modules if and only if
	\begin{equation}\label{equi}
		\iint_{R_i}\sum_{v_j\in V}\rho_j(u,v)\mathrm{d}u \mathrm{d}v=Area(R_i).
	\end{equation}
\end{theorem}

According to Lemma \ref{lem:2} and  Theorem \ref{thm:1}, it is obvious that
\begin{equation*}
	\iint_{R_i}\sum_{v_j\in V}\rho_j(u,v)\mathrm{d}u \mathrm{d}v\ge Area(R_i).
\end{equation*}
Moreover, by Theorem \ref{thm:1}, the non-overlapping constraint in Eq. \eqref{mod:1} is equivalent to Eq. \eqref{equi}.
Therefore, by the regularization technique in machine learning \cite{2014book}, 
the fixed-outline floorplanning problem \eqref{mod:1}  can be transformed to
\begin{equation}\label{mod:non-lap:0}
	\begin{aligned}
		\min \quad & WL(E) + \lambda\sum_{v_i\in V}\iint_{R_i}\sum_{v_j\in V}\rho_j(u,v)\mathrm{d}u \mathrm{d}v \\
		s.t. \quad & \mbox{aspect\ ratio\ is\ suitable}                                      \\
		           & \mbox{all\ modules\ are\ within\ the\ fixed-outline}.
	\end{aligned}
\end{equation}


In the floorplanning model \eqref{mod:non-lap:0},
$\sum_{v_j\in V}\rho_j(u,v)$ is interpreted as the density function within the floorplanning region. According to  VLSI global placement, there are three optional methods to relax the density function for facilitating optimization as follows:

1)  Divide the floorplanning area into bins,  smooth the corresponding density function with the Bell-shaped function, and then make integration;

2)  Divide the floorplanning area into bins, smooth the corresponding density function with the Poisson's equation, and then make integration;

3)  Make integration of $\sum_{v_j\in V}\rho_j(u,v)$ directly.

In this paper, we use scheme 2) since
Poisson's equation method has achieved state-of-the-art results in VLSI global placement.

Let $\psi(u,v)$ be the function given by the Poisson's equation \eqref{eq:poisson}
to smooth the density function $\sum_{v_j\in V}\rho_j(u,v)$ defined in Eq. \eqref{eq:poisson:solver}.
%
%
%
Suggested by  Eq. \eqref{mod:non-lap:0}, for each module $v_i\in V$,
we define the potential energy $N_i$  as
\begin{equation}
	\begin{aligned}\label{eq:poisson:N_i}
		N_i =\iint_{R_i}\psi(u,v)\mathrm{d}u \mathrm{d}v,
	\end{aligned}
\end{equation}
where the integration area $R_i=[x_i-\frac{w_i}{2}, x_i+\frac{w_i}{2}] \times [y_i-\frac{h_i}{2}, y_i+\frac{h_i}{2}]$, $h_i=\frac{A_i}{w_i}$.

Observing  Eq. \eqref{eq:poisson:N_i}, for a soft module,
the potential energy $N_i$ is a function of variables $x_i, y_i, w_i$. While, for a hard module, the potential energy $N_i$ is a function of  variables $x_i$ and $y_i$. Therefore, for convenience we denote
\begin{small}
	\begin{equation}\label{eq:defg}
		\begin{aligned}
			 & g_i(x_i,y_i,w_i)=N_i=\iint_{R_i}\psi(u,v)\mathrm{d}u \mathrm{d}v, ~\forall v_i\in V_s,    \\
			 & g_i(x_i,y_i)=N_i=\iint_{R_i}\psi(u,v)\mathrm{d}u \mathrm{d}v, ~\forall v_i\in V_h\mbox{.}
		\end{aligned}
	\end{equation}
\end{small}

According to  \eqref{eq:defg} and the fixed-outline floorplanning problem  \eqref{mod:non-lap:0}, we can establish the global floorplanning model:
\begin{subequations}\label{mod:3}
	\begin{small}
		\begin{align}
			\min\limits_{\substack{x, y, w}} & \quad LSE(E)                                                                               
			+\lambda[
				\sum_{v_i\in V_s} \!\!g_i(x_i,y_i,w_i) +\! \sum_{v_i\in V_h}\!\! g_i(x_i,y_i)] \label{mod:3:st0}                              \\
			s.t.                             & \quad AR_i^l\le \frac{A_i}{w_i^2}\le AR_i^u, \quad \forall v_i\in V_s \label{mod:3:st1}    \\
			                                 & \quad \frac{w_i}{2} \le x_i\le W-\frac{w_i}{2}, \quad  \forall v_i\in V  \label{mod:3:st2} \\
			                                 & \quad \frac{h_i}{2} \le y_i\le H-\frac{h_i}{2}, \quad  \forall v_i\in V. \label{mod:3:st3}
		\end{align}
	\end{small}
\end{subequations}
In Eq. \eqref{mod:3}, 
the first constraint  ensures  the aspect ratio of each soft module be between the upper and lower bounds.
The last two constraints 
indicate that all modules are located within the floorplanning area.

In order to use  nonlinear optimization method to solve problem \eqref{mod:3},
partial derivative of the objective function with respect to each variable
needs to be calculated.
Since the wirelength function $LSE(E)$ is in analytical form, the partial derivative
is easy to calculate.
Moreover, the difference between the potential energy functions of soft modules and
hard modules is only in the variable $w_i$.
Therefore, we only show calculations of the partial derivatives of
potential energy functions of soft modules in the next subsection.

\subsection{Partial Derivative   of
	Potential Energy}\label{lab:pdc}

In order to calculate the partial derivative of $g_i(x_i,y_i,w_i)$,
first we need to calculate $g_i(x_i,y_i,w_i)$ itself.
According to Section \ref{lab:bsfc}, Poisson's equation has the solution as in Eq. \eqref{eq:poisson:solver}. Substituting it into  Eq. \eqref{eq:defg} can get
\begin{equation}\label{eq:poisson::g}
	\begin{footnotesize}
		\begin{aligned}
			  & g_i(x_i,y_i,w_i)
			=  \iint_{R_i}\psi(z,v)\mathrm{d}z\mathrm{d}v                                         \\
			= & \sum_{u=0}^K\sum_{p=0}^K a_{u,p}
			\int_{x_i-\frac{w_i}{2}}^{x_i+\frac{w_i}{2}}\cos(\frac{u\pi}{W}z)\mathrm{d}z
			\int_{y_i-\frac{h_i}{2}}^{y_i+\frac{h_i}{2}}\cos(\frac{p\pi}{H}v)\mathrm{d}v \\
			= & \sum_{u=1}^K\sum_{p=1}^K a_{u,p}
			\frac{WH}{up\pi^2}
			\left[
				\sin\frac{u\pi(x_i+\frac{w_i}{2})}{W}
				-\sin\frac{u\pi(x_i-\frac{w_i}{2})}{W}
				\right]                                                             \\
			  & \qquad \left[
				\sin\frac{p\pi(y_i+\frac{h_i}{2})}{H}
				-\sin\frac{p\pi(y_i-\frac{h_i}{2})}{H}
				\right]                                                             \\
			  & +\sum_{u=1}^K a_{u,0}
			\frac{W h_i }{u\pi}
			\left[
				\sin\frac{u\pi(x_i+\frac{w_i}{2})}{W}
				-\sin\frac{u\pi(x_i-\frac{w_i}{2})}{W}
				\right]                                                             \\
			  & +\sum_{p=1}^K a_{0,p}
			\frac{H w_i}{p\pi}\left[
				\sin\frac{p\pi(y_i+\frac{h_i}{2})}{H}
				-\sin\frac{p\pi(y_i-\frac{h_i}{2})}{H}
				\right]\mbox{.}
		\end{aligned}
	\end{footnotesize}
\end{equation}
Next, calculate the partial derivatives of \eqref{eq:poisson::g}
with respect to $x_i$, $y_i$ and $w_i$ respectively:
\begin{equation}
	\begin{small}
		\begin{aligned}
			  & \frac{\partial g_i(x_i,y_i,w_i)}{\partial x_i} \\
			= & \sum_{u=1}^K\sum_{p=1}^K a_{u,p}
			\frac{H}{p\pi}
			\left[
				\cos\frac{u\pi(x_i+\frac{w_i}{2})}{W}
				-\cos\frac{u\pi(x_i-\frac{w_i}{2})}{W}
				\right]                                            \\
			  & \qquad \left[
				\sin\frac{p\pi(y_i+\frac{h_i}{2})}{H}
				-\sin\frac{p\pi(y_i-\frac{h_i}{2})}{H}
				\right]                                            \\
			  & +\sum_{u=1}^K a_{u,0}h_i
			\left[
				\cos\frac{u\pi(x_i+\frac{w_i}{2})}{W}
				-\cos\frac{u\pi(x_i-\frac{w_i}{2})}{W}
				\right],                                           \\
		\end{aligned}
	\end{small}
	\label{eq:px}
\end{equation}
\begin{equation}
	\begin{small}
		\begin{aligned}
			  & \frac{\partial g_i(x_i,y_i,w_i)}{\partial y_i} \\
			= & \sum_{u=1}^K\sum_{p=1}^K a_{u,p}
			\frac{W}{u\pi}
			\left[
				\sin\frac{u\pi(x_i+\frac{w_i}{2})}{W}
				-\sin\frac{u\pi(x_i-\frac{w_i}{2})}{W}
				\right]                                            \\
			  & \qquad\left[
				\cos\frac{p\pi(y_i+\frac{h_i}{2})}{H}
				-\cos\frac{p\pi(y_i-\frac{h_i}{2})}{H}
				\right]                                            \\
			  & +\sum_{p=1}^K a_{0,p}
			w_i\left[
				\cos\frac{p\pi(y_i+\frac{h_i}{2})}{H}
				-\cos\frac{p\pi(y_i-\frac{h_i}{2})}{H}
				\right],
		\end{aligned}
	\end{small}
	\label{eq:py}
\end{equation}
\begin{equation}
	\begin{scriptsize}
		\begin{aligned}
			  & \frac{\partial g_i(x_i,y_i,w_i)}{\partial w_i} \\
			= & \sum_{u=1}^K\sum_{p=1}^K a_{u,p}
			\frac{WH}{2up\pi}\!\left\{\!
			\frac{u}{W}\!\!\left[
				\cos\frac{u\pi(x_i+\frac{w_i}{2})}{W}
				+\cos\frac{u\pi(x_i-\frac{w_i}{2})}{W}
				\right]\right.                                     \\
			  & \left.\qquad\qquad\left[
				\sin\frac{p\pi(y_i+\frac{h_i}{2})}{H})
				-\sin\frac{p\pi(y_i-\frac{h_i}{2})}{H}
				\right]\right.                                     \\
			  &                                                
			\left.\qquad
			-\frac{p A_i}{Hw_i^2}
			\left[
				\sin\frac{u\pi(x_i+\frac{w_i}{2})}{W}
				-\sin\frac{u\pi(x_i-\frac{w_i}{2})}{W}
				\right]\right.                                     \\
			  & \left.\qquad\qquad\left[
				\cos\frac{p\pi(y_i+\frac{h_i}{2})}{H}
				+\cos\frac{p\pi(y_i-\frac{h_i}{2})}{H}
				\right]
			\right\}                                           \\
			  & +\sum_{u=1}^K a_{u,0}
			\frac{W}{u\pi}
			\left\{
			\frac{u\pi h_i}{2W}
			\left[
				\cos\frac{u\pi(x_i+\frac{w_i}{2})}{W}
				+\cos\frac{u\pi(x_i-\frac{w_i}{2})}{W}
				\right]\right.                                     \\
			  & \left.\qquad-\frac{A_i}{w_i^2}
			\left[
				\sin\frac{u\pi(x_i+\frac{w_i}{2})}{W}
				-\sin\frac{u\pi(x_i-\frac{w_i}{2})}{W}
				\right]
			\right\}                                           \\
			  & +\sum_{p=1}^K a_{0,p}
			\frac{H}{p\pi}
			\left\{
			\left[
				\sin\frac{p\pi(y_i+\frac{h_i}{2})}{H}
				-\sin\frac{p\pi(y_i^0-\frac{h_i}{2})}{H}
				\right]
			\right.                                            \\
			  & \left.\qquad
			-\frac{p\pi A_i}{2Hw_i}
			\left[
				\cos\frac{p\pi(y_i+\frac{h_i}{2})}{H}
				+\cos\frac{p\pi(y_i-\frac{h_i}{2})}{H}
				\right]\right\}\mbox{.}
		\end{aligned}
	\end{scriptsize}
	\label{eq:pw}
\end{equation}
At this point, the partial derivative calculations are completed.


For each module $v_i$, according to Eqs. \eqref{eq:px}, \eqref{eq:py} and \eqref{eq:pw}, the time complexity required to calculate the partial derivatives with respect to the variables $x_i$, $y_i$ and $w_i$
are $O(K^2)$, when the  coefficients are given. Here
$K$ is the truncation constant in Eq. \eqref{eq:poisson:solver},
which is used to control the accuracy of the solution.
Therefore, the time complexity of calculating partial derivatives
for all $n$ modules is $O(nK^2)$, which
is linear with $n$.
However, experiments show that the constant $K^2$ is very large. Thus
even if the run-time increases linearly with the scale of the problem,
it still takes a large run-time to solve a large-scale problem.
Based on this, next section looks for a way to further reduce the  run-time complexity.

\subsection{Fast Approximate Calculation of Partial Derivatives of
	Potential Energy}\label{lab:sjyh}


From  Eqs. \eqref{eq:p:b} and \eqref{eq:aup:new}, the floorplanning area $[0,W]\times [0,H]$ has been discretized into $K\times K$ bins. 
The width and height of each bin are $w_b=\frac WK$ and $h_b=\frac HK$, respectively.
Note that $R_i$ is the rectangular area occupied by module $v_i$. Let $B_{v_i}$ be  the set of bins occupied by $R_i$, and denote
the right, left, upper, and lower boundaries of $B_{v_i}$
as $X_U^i$, $X_L^i$, $Y_U^i$ and $Y_L^i$, respectively.
That is, $B_{v_i}=\{bin_{p,q}|X_L^i\leq p\leq X_U^i, \ Y_L^i\leq q\leq Y_U^i\}$.

First, we use auxiliary functions to determine the minimum number
of bins covering the rectangular area $R_i$. The auxiliary functions are defined as follows:
\begin{equation}\label{eq:fun:help}
	\begin{aligned}
		up(x)=   & \begin{cases}
			0,                & if\ x=0;   \\
			\lceil x\rceil-1, & else\ x>0,
		\end{cases}         \\
		down(x)= & \lfloor x\rfloor, ~~x\ge 0\mbox{.}
	\end{aligned}
\end{equation}


Then, the four boundaries of the minimum bins covering the rectangular
area $R_i$ can be calculated by:
\begin{equation*}
	\begin{aligned}
		X_L^i= & down( \frac{x_i-\frac{w_i}{2}}{w_b}), &
		X_U^i= & up(\frac{x_i+\frac{w_i}{2}}{w_b});      \\
		Y_L^i= & down(\frac{y_i-\frac{h_i}{2}}{h_b}),  &
		Y_U^i= & up(\frac{y_i+\frac{h_i}{2}}{h_b}).
	\end{aligned}
\end{equation*}

And, the bins enclosed in the rectangle with
lower left corner $(X_L^iw_b,Y_L^ih_b)$
and  upper right corner $((X_U^i+1)w_b$, $(Y_U^i+1)h_b)$
is exactly the set $B_{v_i}$ of bins occupied by module $v_i$. Further, let $\hat{X_U^i}=  X_U^i+1$,   $\hat{Y_U^i}= Y_U^i+1$.

\begin{figure}[h]
	\centering
	\includegraphics[width=7cm]{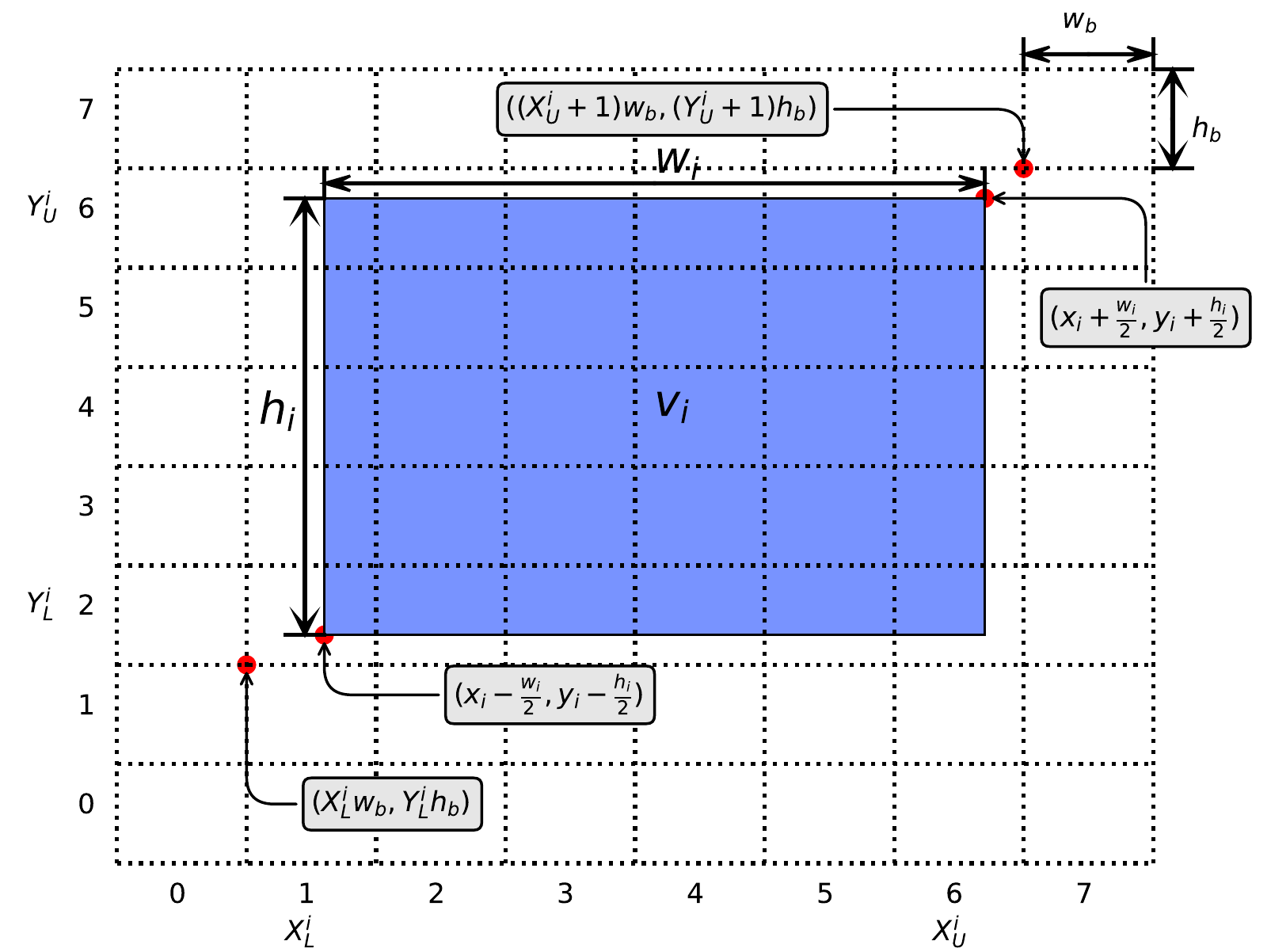}
	\caption{Rectangular area $R_i$ and the squares it occupies.}
	\label{pic:fastbox}
\end{figure}

Fig. \ref{pic:fastbox} shows the specific meaning of the above notations.
Based on the rectangle rule in numerical integration, the integration in Eq. \eqref{eq:defg} can be approximated as
\begin{align*}
\iint_{R_i}\psi(z,v)\mathrm{d}z\mathrm{d}v \approx\sum_{bin_{p.q}\in B_{v_i}} Area(bin_{p,q}\cap R_i) \hat{\psi}(p,q).
\end{align*}
Expanding the right term of the above equation, we get
\begin{align*}\label{eq:fast:g}
	 & g_i(x_i,y_i,w_i)
	\approx\sum_{bin_{p.q}\in B_{v_i}} Area(bin_{p,q}\cap R_i) \hat{\psi}(p,q)                                    \\
	 & = \sum\nolimits_{p=X_L^i}^{X_U^i}\sum\nolimits_{q=Y_L^i}^{Y_U^i}w_bh_b\hat{\psi}(p,q)                      \\
	 & \quad-\sum\nolimits_{q=Y_L^i}^{Y_U^i}h_b [(x_i-\frac{w_i}{2})-X_L^i w_b]\hat{\psi}(X_L^i,q)                \\
	 & \quad-\sum\nolimits_{q=Y_L^i}^{Y_U^i}h_b [\hat{X_U^i} w_b-(x_i+\frac{w_i}{2})]\hat{\psi}(X_U^i,q)          \\
	 & \quad -\sum\nolimits_{p=X_L^i}^{X_U^i}w_b [(y_i-\frac{h_i}{2})-Y_L^i h_b ]\hat{\psi}(p,Y_L^i)              \\
	 & \quad -\sum\nolimits_{p=X_L^i}^{X_U^i} w_b [\hat{Y_U^i} h_b-(y_i+\frac{h_i}{2})]\hat{\psi}(p,Y_U^i)        \\
	 & \quad +[(x_i-\frac{w_i}{2})-X_L^i w_b] [\hat{Y_U^i} h_b-(y_i+\frac{h_i}{2})]\hat{\psi}(X_L^i,Y_U^i)        \\
	 & \quad +[(x_i-\frac{w_i}{2})-X_L^i w_b] [(y_i-\frac{h_i}{2})-Y_L^i h_b ]\hat{\psi}(X_L^i,Y_L^i)             \\
	 & \quad + [\hat{X_U^i} w_b-(x_i+\frac{w_i}{2})] [\hat{Y_U^i} h_b-(y_i+\frac{h_i}{2})]\hat{\psi}(X_U^i,Y_U^i) \\
	 & \quad +[\hat{X_U^i} w_b-(x_i+\frac{w_i}{2})][(y_i-\frac{h_i}{2})-Y_L^i h_b]\hat{\psi}(X_U^i,Y_L^i).
\end{align*}

In the above equation, $w_i$ appears as a variable explicitly. Moreover, $h_i=\frac{A_i}{w_i}$. Hence, by taking the partial derivative of the last equality with respect to $w_i$, we get

\begin{equation}\label{eq:fast:pw}
	\begin{aligned}
		 & \frac{\partial g_i(x_i,y_i,w_i)}{\partial w_i}
		\approx \sum\nolimits_{q=Y_L^i}^{Y_U^i}\frac{h_b}{2} [\hat{\psi}(X_L^i,q)+\hat{\psi}(X_U^i,q)]                                             \\
		 & \quad  -\sum\nolimits_{p=X_L^i}^{X_U^i}\frac{A_iw_b}{2w_i^2} [\hat{\psi}(p,Y_L^i)+\hat{\psi}(p,Y_U^i)]                                  \\
		 & \quad-\frac{1}{2} [(y_i-\frac{h_i}{2})-Y_L^i h_b] [\hat{\psi}(X_L^i,Y_L^i)+\hat{\psi}(X_U^i,Y_L^i)]                                     \\
		 & \quad -\frac{1}{2}[\hat{Y_U^i} h_b-(y_i+\frac{h_i}{2})] [\hat{\psi}(X_L^i,Y_U^i)+\hat{\psi}(X_U^i,Y_U^i)]                               \\
		 & \!\!\!\!\!\!\!\!\!\! \quad +\frac{A_i}{2w_i^2}[(x_i-\frac{w_i}{2})-X_L^i w_b] [\hat{\psi}(X_L^i,Y_L^i)+\hat{\psi}(X_L^i,Y_U^i)]         \\
		 & \!\!\!\!\!\!\!\!\!\! \quad +\frac{A_i}{2w_i^2} [\hat{X_U^i} w_b-(x_i+\frac{w_i}{2})] [\hat{\psi}(X_U^i,Y_L^i)+\hat{\psi}(X_U^i,Y_U^i)].
	\end{aligned}
\end{equation}

Similarly, the partial derivatives of $g_i(x_i,y_i,w_i)$ with respect to $x_i$ and $y_i$ can be approximated by:

\begin{equation}\label{eq:fast:px}
	\begin{aligned}
		 & \frac{\partial g_i(x_i,y_i,w_i)}{\partial x_i}
		\approx
		\sum_{q=Y_L^i}^{Y_U^i}
		h_b\left[-\hat{\psi}(X_L^i,q)+\hat{\psi}(X_U^i,q)\right]                                                                    \\
		 & \! \quad +\!\left[(y_i-\frac{h_i}{2})-Y_L^i h_b\right]\left[\hat{\psi}(X_L^i,Y_L^i)-\hat{\psi}(X_U^i,Y_L^i)\right]       \\
		 & \!\quad +\!\left[\hat{Y_U^i} h_b-(y_i+\frac{h_i}{2})\right]\left[\hat{\psi}(X_L^i,Y_U^i)-\hat{\psi}(X_U^i,Y_U^i)\right],
	\end{aligned}
\end{equation}

\begin{equation}\label{eq:fast:py}
	\begin{aligned}
		 & \frac{\partial g_i(x_i,y_i,w_i)}{\partial y_i}
		\approx
		\sum_{p=X_L^i}^{X_U^i}
		w_b\left[-\hat{\psi}(p,Y_L^i)+\hat{\psi}(p,Y_U^i)\right]                                                                        \\
		 & \!\quad +\!\left[(x_i-\frac{w_i}{2})-X_L^i w_b\right]\!\left[\hat{\psi}(X_L^i,Y_L^i)-\hat{\psi}(X_L^i,Y_U^i)\right]          \\
		 & \!\quad +\!\left[\hat{X_U^i} w_b-(x_i+\frac{w_i}{2})\right]\!\!\left[\hat{\psi}(X_U^i,Y_L^i)-\hat{\psi}(X_U^i,Y_U^i)\right].
	\end{aligned}
\end{equation}

According to Eqs. \eqref{eq:fast:pw}, \eqref{eq:fast:px} and \eqref{eq:fast:py}, 
the computing cost for $\frac{\partial g_i(x_i,y_i,w_i)}{\partial x_i}$,
$\frac{\partial g_i(x_i,y_i,w_i)}{\partial y_i}$ and
$\frac{\partial g_i(x_i,y_i,w_i)}{\partial w_i}$
is a linear function of $(X_U^i-X_L^i)+(Y_U^i-Y_L^i)$, if all $\hat{\psi}(p, q)$ have been computed.
Therefore, the time complexity of calculating the partial derivatives  for all modules is $O(n \frac {AVG}n)$, where $AVG=\sum_{i=1}^n (X_U^i-X_L^i+Y_U^i-Y_L^i)$, if all $\hat{\psi}(p, q)$ have been computed.

According to Eq. \eqref{eq:aup:new}, we first invoke the Fast Fourier Transform (FFT) one time to calculate all the coefficients $a_{up}$, and then invoke the FFT another time to calculate all values of $\hat{\psi}(p, q)$. Since the time complexity of one FFT is $O(K^2\log K)$,  all the values  $\hat{\psi}(p, q)$ can be calculated in time $O(K^2\log K)$ by the FFT on Eq. \eqref{eq:aup:new}. Hence, the total time complexity of calculating the partial derivatives for all modules  is  $O(K^2\log K)+ O(n \frac {AVG}n)$.


It must be remarked	that, in ePlace \cite{2015ePlace} each iteration requires invoking the FFT  three times for  calculating  the partial derivatives for all modules (see Eqs. (21) and (24) in \cite{2015ePlace}).
Further, in our fast approximate calculation  of partial derivatives, it is apparently that the average number $\frac {AVG}n $
of squares occupied by all modules is much smaller than the total number $K^2$ of squares. So the time complexity of this fast approximate calculation of partial derivatives still has a linear relationship
with the total number of modules, but the difference is that the constant $\frac {AVG} n$ is no longer related to
the truncation constant $K^2$ as in Eqs. \eqref{eq:px}, \eqref{eq:py} and \eqref{eq:pw}. Therefore, the  cost of calculating all partial derivatives is reduced.

For the convenience of description, we will call the method in Section \ref{lab:pdc} the partial derivative calculation method before acceleration (PDBA), and the method in this section the accelerated partial derivative calculation method.

\section{Fixed-outline Floorplanning Algorithm}\label{lab:sf}
\begin{figure}[h]
	\centering
	\includegraphics[width=1.0\hsize]{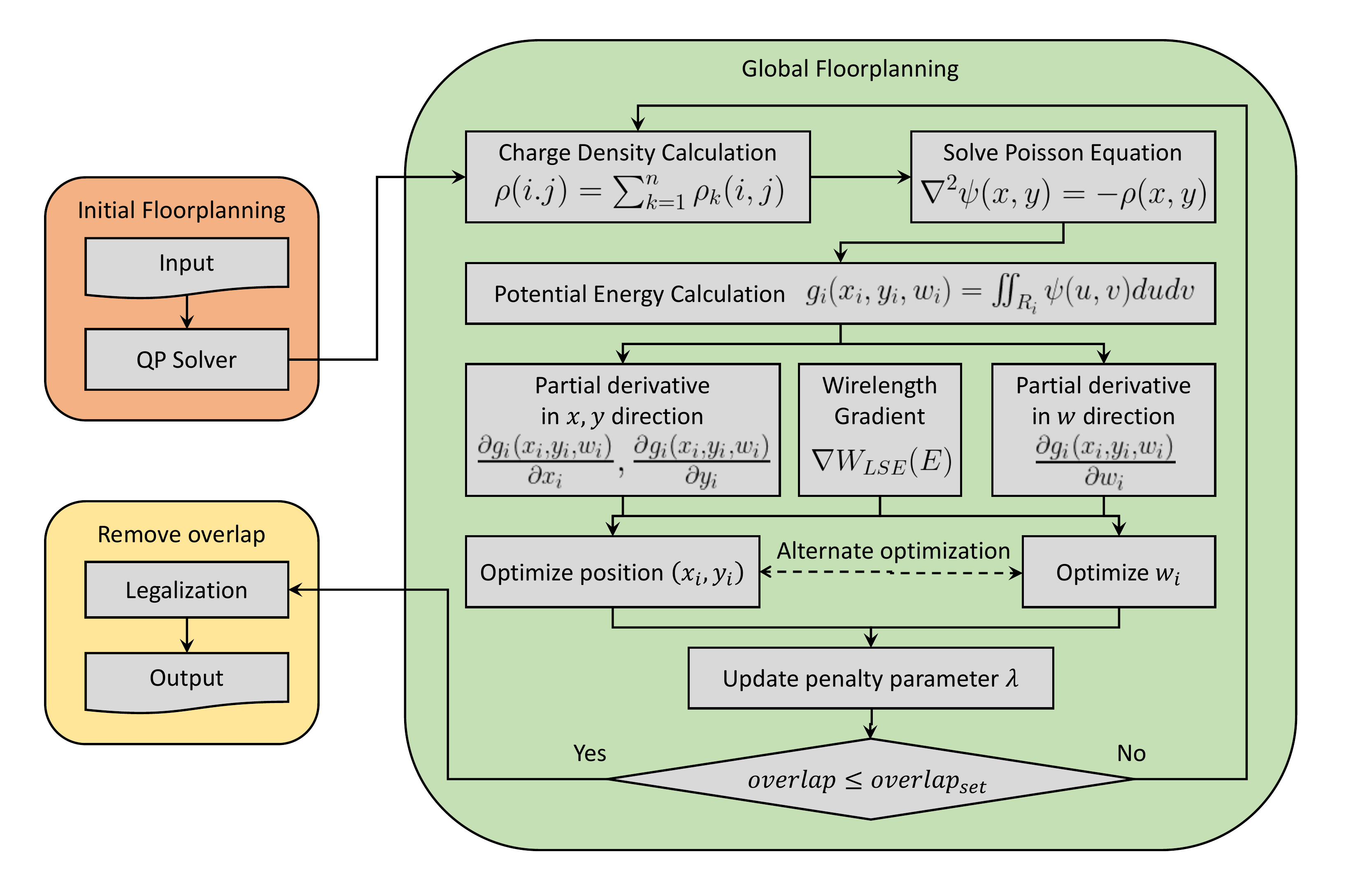}
	\caption{Framework of our fixed-outline floorplanning algorithm.}
	\label{pic:allf}
\end{figure}

In this section, we present our algorithm for fixed-outline floorplanning,
as shown in Fig. \ref{pic:allf}. It contains three phases: initial floorplanning, global floorplanning and legalization.
We directly adopts the QP method in ePlace \cite{2015ePlace} to produce an initial floorplan.
Hence, we  mainly introduce in this section the global floorplanning
and legalization algorithms.

\subsection{Global Floorplanning Algorithm}
In the global floorplanning phase of Fig. \ref{pic:allf},
Charge Density Calculation, Solve Poisson's Equation, Potential Energy Calculation, Partial derivative in $x,y$ directions, Partial derivative in $w$ direction	and Wirelength Gradient have been explained in Section \ref{lab:mxjlyqj}.
%
In this subsection, we present other parts of the global floorplanning algorithm for problem \eqref{mod:3}, as in Algorithm \ref{alg:n:all}.

\begin{algorithm}[h]
	\caption{Global floorplanning algorithm} \label{alg:n:all}
	\begin{algorithmic}[1]
		\small
		\Require initial solution $u_0=(x_0,y_0,w_0)$, $\lambda_0$, maximum iterations $k_{max}$, minimum overlap $O_{min}$;
		\Ensure solution $u_{gp}$.
		\For{$k=0\rightarrow k_{max}$}
		\State \!\!\! $f_{k}=LSE(E)+\lambda_k [\sum_{v_i\in V_s} g_i(x_i,y_i,w_i) + \sum_{v_i\in V_h}g_i(x_i,$ $y_i)$];
			\State calculate partial derivatives
		$\nabla_x f_k$, $\nabla_y f_k$ and $\nabla_w f_k$ of $f_{k}$;
			\State $(x_{k+1},y_{k+1})=$Nesterov-Solver$((x_{k},y_{k}),f_k)$;
			\State $w_{k+1}=$Normal-Solver$(w_k,1,f_k)$;
			\State $u_{gp}=u_{k+1}=(x_{k+1},y_{k+1},w_{k+1})$;
			\State project $u_{k+1}$ onto the feasible region;
			\State calculate overlap $O_{k+1}$ and update $\lambda_{k}$ to $\lambda_{k+1}$;
			\If{$O_{k+1}\le O_{min}$}
			\State break;
			\EndIf
			\EndFor
			\State \Return $u_{gp}$.
	\end{algorithmic}
\end{algorithm}

In Algorithm \ref{alg:n:all}, we use projected gradient method to solve problem \eqref{mod:3}. Line 2 
constructs the objective function $f_k$ required in the $k$-th iteration. Line 3 calculates the partial derivative of the objective function in each direction, which will be used in lines 4-5.

We use Nesterov's method \cite{2015ePlace} to optimize the module positions $(x_i, y_i)$. In line 4 of Algorithm \ref{alg:n:all}, we first fix the widths of soft modules and call Nesterov's method to optimize module positions (Algorithm \ref{alg:ns}), which was first adopted for VLSI placement in ePlace \cite{2015ePlace}.
Nesterov's method is the first order optimization algorithm with  convergence rate $O(1/k^2)$, where $k$ is the number of iterations.
We set $a_0=1$, $u_0=v_0$ and $v_0$ is the solution given by the initial floorplanning.

\begin{algorithm}[h]
	\caption{Nesterov-Solver($u_k$,$f_k$) //Nesterov's method at the $k$-th iteration \cite{2015ePlace}} \label{alg:ns}
	\hspace*{0.02in} {\bf Input:} 
	major solution $u_k$, reference solution $v_k$,
	optimization parameter $a_k$ and objective function $f_k=f(v_k)$;\\
	\hspace*{0.02in} {\bf Output:} 
	$u_{k+1}$, $v_{k+1}$.
	\begin{algorithmic}[1]
		\State calculate the gradient in the $x,y$ direction: $\nabla_{x,y} f_k=\nabla_{x,y} f_k(v_k)$;
		\State steplength $a_k=\mathop{\arg\!\max}\limits_{a}\{f_k-f(v_k-a\nabla_{x,y} f_k)\ge 0.5a\|\nabla_{x,y} f_k\|^2\}$;
		\State new solution $u_{k+1}=v_k-a_k\nabla_{x,y} f_k$;
		\State parameter update $a_{k+1}=\frac{1+\sqrt{4a_k^2+1}}{2}$;
		\State new reference solution $v_{k+1}=u_{k+1}+\frac{(a_k-1)(u_{k+1}-u_{k})}{a_{k+1}}$;
		\State \Return $u_{k+1}$, $v_{k+1}$.
	\end{algorithmic}
\end{algorithm}


In Nesterov's method, the same as ePlace \cite{2015ePlace}, we use Jacobi preconditioner to get faster convergence rate and better solution.
In Nesterov's method, since we have fixed the widths of soft modules when we call the Nesterov-Solver, we use the same diagonal Jacobi preconditioner as ePlace \cite{2015ePlace} to get faster convergence and better solution.

Line 5 of Algorithm \ref{alg:n:all} uses normalized gradient descent method
to optimize the widths of soft modules.
By fixing the positions of modules obtained in line 4, we calculate
$\nabla_w f_k=\nabla_w f_k(u_k)$. Then, we set  $u_{k+1}=u_k-step\nabla_w f_k$, 
in which $step=a/\|\nabla_w f_k\|$, and here we set the constant $a=1$.
Note that, after optimizing the widths $w_i$ of soft modules,  the positions of pins will be affected. Hence, in order to ensure the optimization accuracy, 
the positions of pins will be updated  proportionally before the next time we call the Nesterov-Solver.

Since $u_{k+1}$ may not satisfy the constraints in problem \eqref{mod:3}.
In order to obtain a feasible solution, line 7 of Algorithm \ref{alg:n:all} projects $u_{k+1}$ onto the feasible region of  problem \eqref{mod:3}, such that it satisfies the constraints \eqref{mod:3:st1}, \eqref{mod:3:st2} and \eqref{mod:3:st3}.
First, to meet the constraint \eqref{mod:3:st1}, line 7 projects the widths $w_i$ of soft modules in $u_{k+1}$ as follows:
\begin{equation*}
	w_i=\begin{cases}\label{eq:lg:w}
		\sqrt{
		\frac{A_i}{AR_L^i}},       & if ~~ \frac{A_i}{w_i^2}<AR_L^i; \\
		\sqrt{\frac{A_i}{AR_U^i}}, & if ~~ \frac{A_i}{w_i^2}>AR_U^i; \\
		w_i,                       & else.                           \\
	\end{cases}
\end{equation*}
Next, line 7 of Algorithm \ref{alg:n:all} projects in the same way the positions of modules such that they satisfy
the constraints \eqref{mod:3:st2} and \eqref{mod:3:st3}:
\begin{equation*}\label{eq:lg:x_y}
	\begin{aligned}
		x_i & =\begin{cases}
			\frac{w_i}{2},   & if ~~x_i<\frac{w_i}{2};   \\
			W-\frac{w_i}{2}, & if ~~x_i>W-\frac{w_i}{2}; \\
			x_i,             & else,
		\end{cases} \\
		y_i & =\begin{cases}
			\frac{h_i}{2},   & if ~~y_i<\frac{h_i}{2};   \\
			H-\frac{h_i}{2}, & if ~~y_i>H-\frac{h_i}{2}; \\
			y_i,             & else\mbox{.}
		\end{cases}
	\end{aligned}
\end{equation*}

Line 8 of Algorithm \ref{alg:n:all} updates the value of penalty parameter $\lambda$
and calculates the overlaps between modules.
Here, the method for updating  the penalty parameter value
is the same as that in ePlace \cite{2015ePlace}.
First, set an expected increase of wirelength  $\Delta HPWL_{ref}$ for each iteration. Then the penalty parameter is updated as
\begin{equation*}
	\lambda_{k+1}=\mu_k \lambda_k,\quad
	\mu_k=\mu_0^{-\frac{\Delta HPWL_k}{\Delta HPWL_{ref}}+1},
\end{equation*}
where $\Delta HPWL_k=HPWL(v_k)-HPWL(v_{k-1})$. 
The constant $\mu_0$ is set to 1.1,
and $\mu_k$ is usually truncated to the interval $[0.75,1.1]$.

Finally, in lines 9-10, Algorithm \ref{alg:n:all} will stop if the current total overlap between modules is less than the threshold $O_{min}$, which
ensures that the overlapping area must be less than $O_{min}$ before legalization begins.

\subsection{Legalization}\label{lab:hfh}

After global floorplanning, we can get a ``rough" floorplanning of modules with a small overlap. However, there is still a little overlap in
the floorplan (more details can be found  in the experimental results section). In other words, it does not satisfy the constraint that there is no overlap between modules. Hence legalization will be used to make the global floorplan be a legal floorplan. Therefore, this section will discuss how to eliminate the overlap between modules.

In legalization, we use the global floorplanning result as the initial solution, and treat all  soft modules as hard ones. If a floorplan is  small-scale, then it is legalized using the $pl2sp$ function in Parquet \cite{2003Fixed} directly; otherwise it is legalized using the method proposed in this subsection.

Abacus \cite{2008Abacus}, SAINT \cite{2016SAINT} and Floorist \cite{2008FLOORIST} are state-of-the-art legalization methods to eliminate overlaps between modules. Abacus\cite{2008Abacus} is used in VLSI placement, in which standard cells can be aligned to  rows. However, in fixed-outline floorplanning, the modules are of various heights.
Hence it is difficult to extend Abacus to legalize a floorplan.	SAINT \cite{2016SAINT} uses polygon to approximate the curve of area of every soft module, and solve an ILP for legalization, which is time-consuming and is difficult to scale up for large-scale floorplanning problem.
Floorist\cite{2008FLOORIST} is a floorplan repair method, its
main idea is as follows. First  a non-overlapping  floorplan is constructed based on the global floorplanning result. If it exceeds the floorplanning area, then the positional relationship of modules is  modified
and the floorplan is compressed, until all modules are located legally in the floorplanning area.

We propose a new legalization algorithm based on Floorist \cite{2008FLOORIST}. 
The biggest difference between our legalization algorithm  and Floorist is as follows: Floorist considers how to insert constraints into the constraint graphs so that all modules finally are non-overlapping,  but does not delete constraints in the constraint graphs. During insertion, the fixed-outline is satisfied but the non-overlapping condition may be violated. A major issue is that the constraint graphs may become very dense finally. In contrast, our algorithm not only inserts constraints but also deletes constraints in the constraint graphs. Moreover, our algorithm  ensures that
the non-overlapping condition is satisfied but the fixed-outline may be violated. Eventually, our algorithm will satisfy the fixed-outline condition by modifying the constraint graphs.

\subsubsection{Horizontal and Vertical Constraint Graphs and
	Related Definitions}\label{lega_def}

First, we construct  horizontal and vertical constraint graphs $HCG$ and  $VCG$, according to the positional relationship of the modules after global floorplanning.
There are two positional cases between  lower left corners of two modules. We
take one case to illustrate how to insert directed edges between two modules in  $HCG$ and  $VCG$.

Assume that the coordinate of the lower left corner of module $A$ is not greater than that of $B$.
Then we have the following processing for modules $A$ and $B$:

(a) If $A$ and $B$ do not overlap both horizontally and vertically, then we have $A\uparrow B \in VCG$ and $A\rightarrow B \in HCG$.

(b) If $A$ and $B$ overlap vertically but do not overlap horizontally, then we have $A\rightarrow B \in HCG$.

(c) Assume that $A$ and $B$ overlap both horizontally and vertically. If the horizontal overlap is greater than the vertical overlap. Then there is $A\uparrow B \in VCG$, otherwise $A\rightarrow B \in HCG$.
\begin{figure}[h]
	\centering
	\subfloat[]{\includegraphics[width=2.5cm]{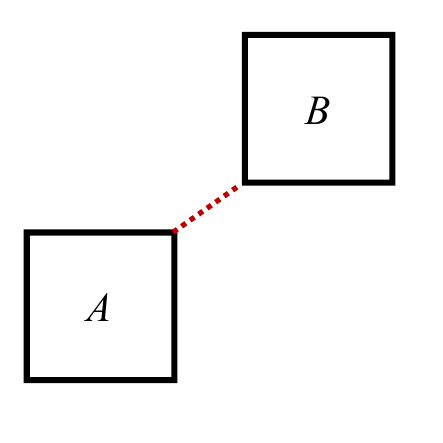}}
	~~
	\subfloat[]{\includegraphics[width=2.5cm]{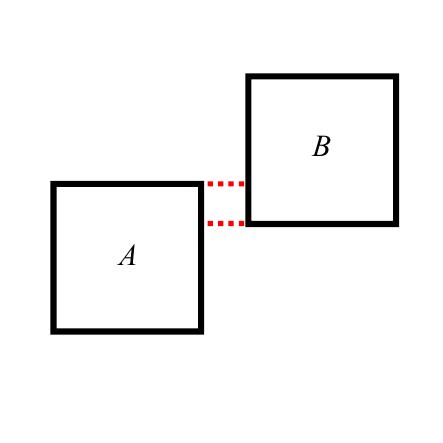}}
	~~
	\subfloat[]{\includegraphics[width=2.5cm]{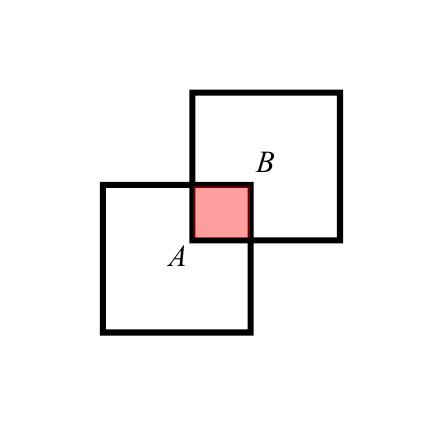}}
	\caption{	
Three overlapping relationships between modules which are used to construct  VCG and HCG.}
	\label{pic:TCG}
\end{figure}

Basically, our algorithm legalizes a floorplan alternatively in the horizontal and vertical directions. In the horizontal direction, our legalization algorithm modifies a floorplan by deleting an edge between two modules in $HCG$, then adding  an edge between the two modules in $VCG$, and compress the floorplan. The vertical legalization operates in a similar way.
Generally, we can arbitrarily delete or add an edge between two modules, if there is no contradiction. Moreover, it is not needed to ensure that $HCG$ and $VCG$ are transitive, for effectively reducing the number of edges in $HCG$ and $VCG$.

However, choosing an edge for adjustment is an important step that affects the effectiveness of our legalization algorithm. During horizontal legalization, we tend to select an edge that has the least impact on $VCG$,  which is achieved by comparing the weights of the edges in $HCG$ defined in Eq. \eqref{eq:hfh:cal4}.  The vertical legalization selects an edge in $VCG$ similarly. The specific details will be given in Algorithm \ref{alg:lg:part}.

During horizontal legalization, according to the $HCG$,
we first move the modules as far as possible to the left in the floorplanning area,  while keeping the $HCG$ unchanged. The method is, we calculate
the abscissa $x_A$ of the lower left corner of every module $A$:
\begin{equation}
	\label{eq:hfh:cal1}
	\begin{aligned}
		x_A= &
		\begin{cases}
			\max\limits_{\forall B(B\rightarrow A)}  \left(x_B+w_B\right), & if\ \exists B(B\rightarrow A); \\
			0,                                                             & else\mbox{.}
		\end{cases}
	\end{aligned}
\end{equation}
Similar to Eq. \eqref{eq:hfh:cal1},
the ordinate $y_A$ of the lower left corner of every module $A$ can be calculated.
After the above calculations, the coordinate $(x_A, y_A)$ of the lower left corner of each module is determined, and all overlaps between modules are eliminated.

Next, let $noRight	=\{A| \forall B, A\not\rightarrow B\}$,
we calculate the width $W_d$ of the smallest rectangle enclosing all modules:
\begin{equation}
	\label{eq:hfh:cal2}
	\begin{aligned}
		W_d= \max_{A\in noRight} \left(x_A+w_A\right).
	\end{aligned}
\end{equation}
Similar to Eq. \eqref{eq:hfh:cal2},
we can calculate the height $H_d$ of the smallest rectangle enclosing all modules.

Then, we change temporarily the origin of the coordinate system to $(W_d, 0)$, and change the positive direction of the $x$-axis to left.
In the temporary coordinate system, we construct the horizontal constraint graph $HCG^{'}$ of the floorplan. According to the $HCG{'}$, by calculating Eq. \eqref{eq:hfh:cal1}, the  lower left corner coordinate $x_A'$ of each module $A$ can be obtained.
After the above calculations, the horizontal slack $S_A^x$ of a module $A$ is defined as

\begin{small}
\begin{equation}
	\label{eq:hfh:cal3}
	S_A^x=W-(x_A+x_A'+w_A),
\end{equation}
\end{small}
here $W$ is the width of the floorplanning area.

Similarly, we can define the vertical slack $S_A^y$ of  module $A$.
Then, we say a horizontal relationship $A\rightarrow B$ is critical if
$S^x_A=0$ and $S^x_B=0$. 
Similarly, we say a vertical relationship $A\uparrow B$ is critical if
$S^y_A=0$ and $S^y_B=0$.

Next, we define the compressible relationship between modules.
Modules $A$ and $B$ has a  compressible relationship in the horizontal direction, denoted as $A\cap_yB=\emptyset$, if $A\rightarrow B$ and
$\left[y_A, y_A+h_A \right]\cap\left[y_B, y_B+h_B \right]=\emptyset$. This definition means that, if there are two modules $A$ and $B$ such that $A\cap_yB=\emptyset$,
then $A$ or $B$ may be moved horizontally such that the floorplan is more compact. Similarly, we can define the vertical compressible relationship, and denote it as $A\cap_xB=\emptyset$. Moreover, for a horizontal critical relationship $A \rightarrow B$, we define  the weight of $A \rightarrow B$ as
\begin{equation}
		\label{eq:hfh:cal4}
		\begin{aligned}
			Weight(A \rightarrow B)= &
			\begin{cases}
				S^y_A-h_B, & if\ y_A\leq y_B; \\
				S^y_B-h_A, & else.
			\end{cases}
		\end{aligned}
	\end{equation}

For $y_A\leq y_B$, if we delete the horizontal critical relationship $A\rightarrow B$ in $HCG$ and add the vertical relationship $A\uparrow B$ in $VCG$,
then the vertical slack of the module $A$ is $S^y_A-h_B$.
Similarly, for $y_A> y_B$, after deleting $A\rightarrow B$ in $HCG$ and adding $B\uparrow A$ in $VCG$, the vertical slack of the module $B$ is about $S^y_B-h_A$.
By setting this weight, when we need to delete an edge in the $HCG$ and add a new edge in the $VCG$, the algorithm will preferentially select the edge with the smallest increase of the height $H_d$ of the smallest rectangle enclosing all modules.
Similarly, we can define the weight of a vertical critical relationship in the vertical constraint graph.

\subsubsection{Legalization Algorithm}

Algorithm \ref{alg:lg:all} shows the overall process of our legalization algorithm, which finds a legalized floorplan within a given number of iterations.
\begin{algorithm}[h]
	\caption{Legalization algorithm}
	\label{alg:lg:all}
	\begin{algorithmic}[1]
		\small
		\Require width and height of floorplanning area $(W,H)$,
		horizontal and vertical constraint graphs $(HCG, VCG)$,
		maximum number of iterations $N$;
		\Ensure 
		coordinate $(\hat{x}_A, \hat{y}_A)$ of lower left corner of each module.
		\For{$i=1\rightarrow N$}
		\State $(HCG, VCG)=LG_x(W,HCG,VCG)$;
		\State $(HCG, VCG)=LG_y(H,HCG,VCG)$;
		\State calculate the smallest rectangle $(W_d, H_d)$
		enclosing all modules according to $(HCG, VCG)$;
		\If{$W_d\leq W$ and $H_d\leq H$}
		\State break;
		\EndIf
		\EndFor
		\State \Return
		$(\hat{x}_A$, $\hat{y}_A)$.
	\end{algorithmic}
\end{algorithm}

Lines 2-3 of Algorithm \ref{alg:lg:all} legalize the floorplan in
the horizontal and vertical directions alternatively, which will be introduced in Algorithm \ref{alg:lg:part}. According to the current $HCG$ and $VCG$, line 4 uses Eq. \eqref{eq:hfh:cal1} to calculate the coordinate $(\hat{x}_A, \hat{y}_A)$ of the lower left corner  of each module,
and uses Eq. \eqref{eq:hfh:cal2} to calculate the smallest rectangle enclosing these modules.
Line 5 of Algorithm \ref{alg:lg:all} determines whether
the smallest rectangle is within the floorplanning area.
If so, exit the iteration.

Algorithm \ref{alg:lg:part} presents our legalization algorithm  in the horizontal  direction. The vertical legalization algorithm is similar, 
and will not be given here.
\begin{algorithm}[h]
	\caption{ $LG_x(W,HCG,VCG)$  // Horizontal legalization} 
	\label{alg:lg:part}
	\begin{algorithmic}[1]
		\small
		\Require width $W$ of floorplanning area,
		horizontal and vertical constraint graphs $(HCG, VCG)$;
		\Ensure $(HCG, VCG)$ after horizontal legalization.
		\State calculate  $(x_A, y_A)$ and $(S^x_A, S^y_A)$ of all modules and the width $W_d$ based on the current $(HCG, VCG)$;
		\While{$W_d > W$}
		\State
		let $S=\{A\rightarrow B\in HCG|S^x_A=0 \  and\   S^x_B=0\}$ be the set of  horizontal critical relationships;
		\If{$\exists (A\rightarrow B)\in S$ and $A\cap_y B=\emptyset$} 
		\State delete  relationship $A\rightarrow B$ from the current $HCG$;
		\Else
		\For{$\forall (A\rightarrow B)\in S$}
		\State calculate
		$
			Weight(A \rightarrow B);
		$
		\EndFor
		\State find  modules $A$ and $B$ with  largest 
			$Weight(A \rightarrow B)$;
		\State delete relationship $A\rightarrow B$ from the current $HCG$;
		\If{$y_A\leq y_B$}
		\State add relationship $A\uparrow B$ to the current  $VCG$;
		\Else
		\State add relationship $B\uparrow A$ to the current $VCG$;
		\EndIf
		\EndIf
		\State do line 1 again;
		\EndWhile
		\State \Return $(HCG, VCG)$.
	\end{algorithmic}
\end{algorithm}

According to the current $HCG$ and $VCG$, lines 1 and 15 use Eq. \eqref{eq:hfh:cal1} to move all modules to the left and bottom in the floorplanning area, and make them non-overlapping.
Then, lines 1 and 15  use Eq. \eqref{eq:hfh:cal3} and Eq. \eqref{eq:hfh:cal2} to
calculate horizontal and vertical slacks $(S^x_A, S^y_A)$ of all modules, and the width $W_d$ of the smallest rectangle enclosing these modules, respectively.

Line 2 determines whether the floorplan obtained from lines 1 or 15 is already in the floorplanning area.
If so, a feasible solution in the horizontal direction has been obtained, and the algorithm stops.
Line 3 of Algorithm \ref{alg:lg:part} finds all critical relationships in the $HCG$. Lines 4-5 determine whether there is a compressible relationship $A\rightarrow B$ in
the set of horizontal critical relationships.
If so, it indicates that  module $B$ can be moved left so that $A$ and $B$ still do not overlap. Therefore, we delete relationship $A\rightarrow B$ from the current $HCG$.

Lines 7-14 deal with the situation that the floorplan cannot be compressed directly in the horizontal direction. In this situation, 
we  use Eq. \eqref{eq:hfh:cal4} to calculate $Weight(A \rightarrow B)$ of all horizontal critical relationships, and find the largest one between lines 7 and 9. Then we delete the $A \rightarrow B$ corresponding to the largest $Weight(A \rightarrow B)$ in $HCG$ at line 10, and finally add a new corresponding $VCG$ edge between lines  11-14. The purpose of this is to minimize the increase in the height $H_d$ of the floorplan after changing the relationship.


Let $HCG$ be $G(V,E)$. According to the topological-sort and
Eqs. \eqref{eq:hfh:cal1}, \eqref{eq:hfh:cal2} and \eqref{eq:hfh:cal3},
the time complexity of lines 1 and 15 in Algorithm \ref{alg:lg:part} is $O(|E|+|V|)$. Note that, $|S|\leq |E|$ holds for the set of horizontal critical relationships.
So, the time complexity of lines 4-5, 7-8 and 9  of Algorithm \ref{alg:lg:part} are all $O(|E|)$. Therefore, the time complexity of each iteration of Algorithm \ref{alg:lg:part} is $O(|E|+|V|)$.

\subsubsection{Example of horizontal legalization}
\begin{figure}[ht]
\subfloat[]{\includegraphics[width=7cm]{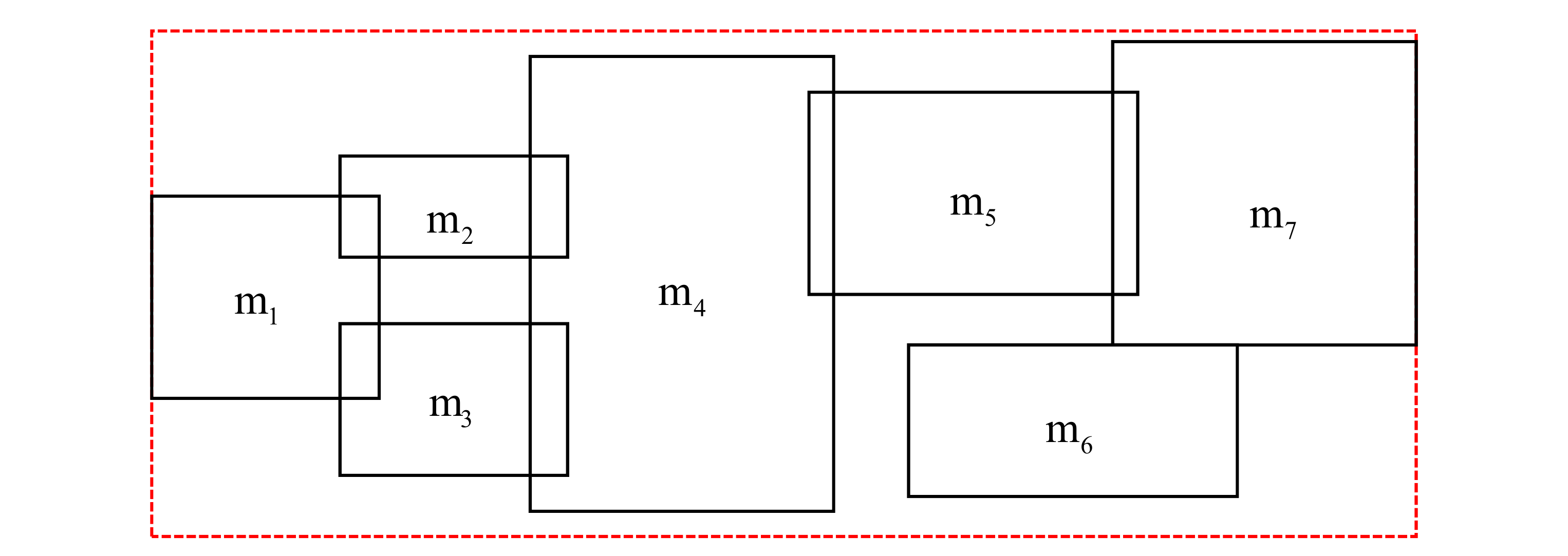}}
	\newline
	\subfloat[]{\includegraphics[width=7cm]{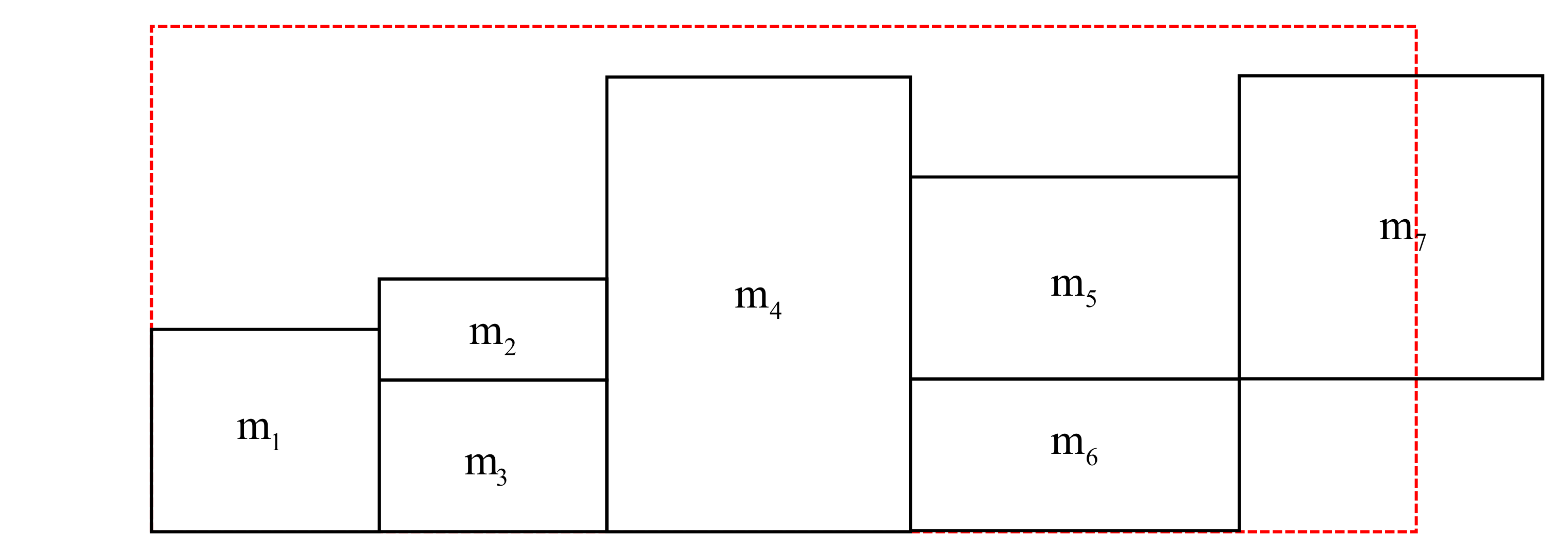}}
	\newline
	\subfloat[]{\includegraphics[width=7cm]{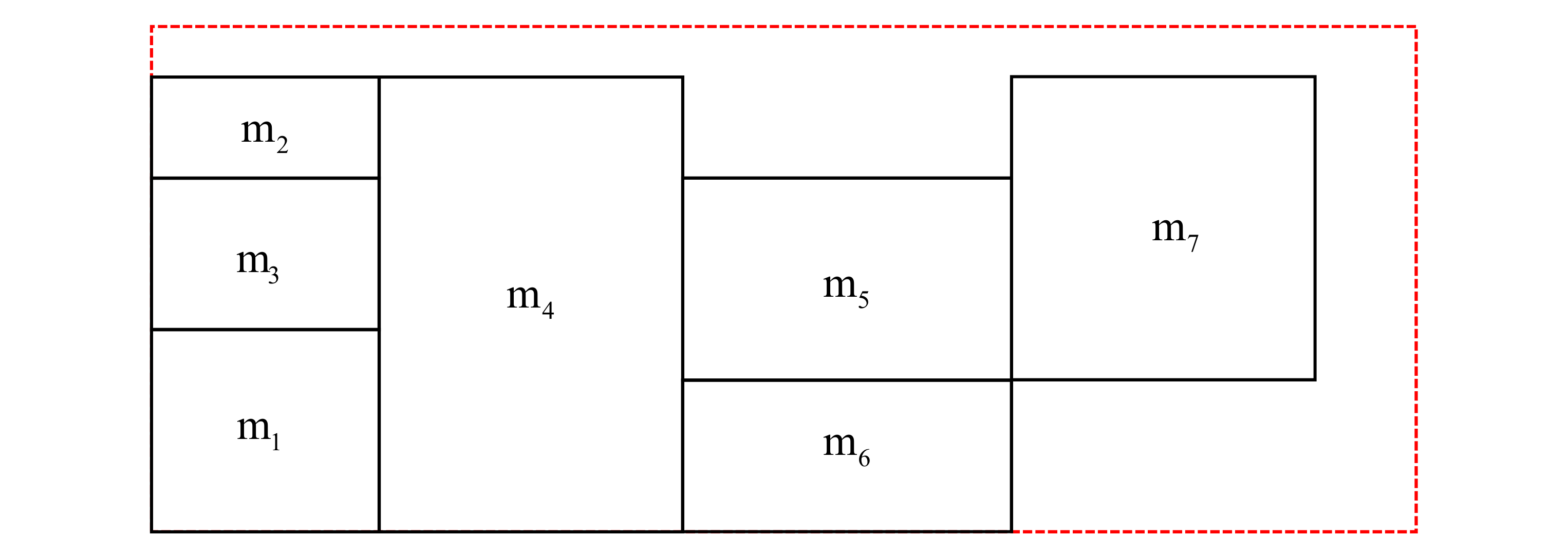}}
	\caption{
		Distribution of modules: (a) modules with overlaps \cite{2008Cong}; (b) floorplan after compression; (c) legal floorplan after horizontal legalization by Algorithm \autoref{alg:lg:part}.}
		\label{pic:leg_exmpl}
\end{figure}

To illustrate the effectiveness of our legalization algorithm, we implement it on the example of XDP \cite{2008Cong} as shown in Fig \ref{pic:leg_exmpl}(a), which was used in  \cite{2008Cong} for mixed-size placement to show that it is nontrivial to identify a right set of edges for adjustment.

\begin{figure}[ht]
	\centering
	~~	\subfloat[]{\includegraphics[width=6cm]{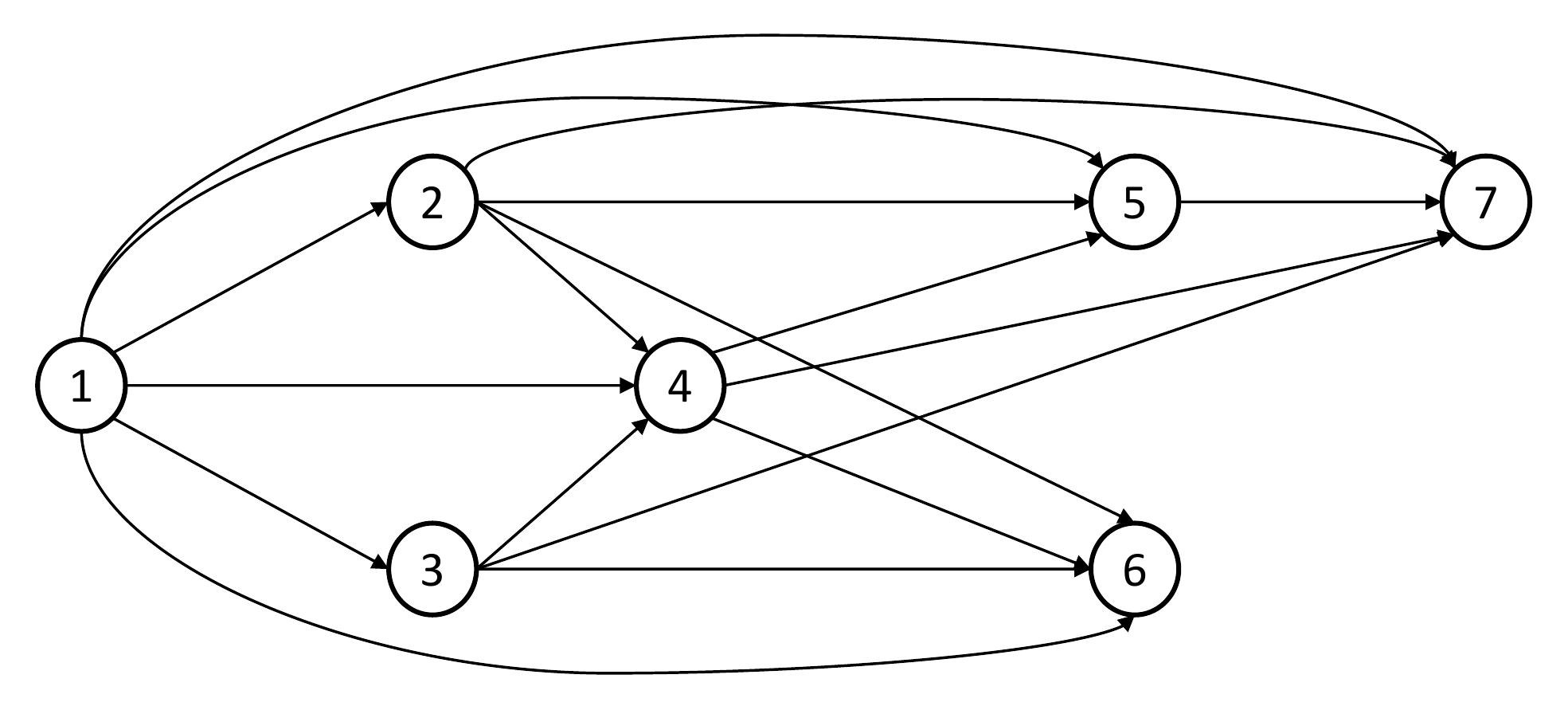}}
	\newline
	\centering
	\subfloat[]{\includegraphics[width=6cm]{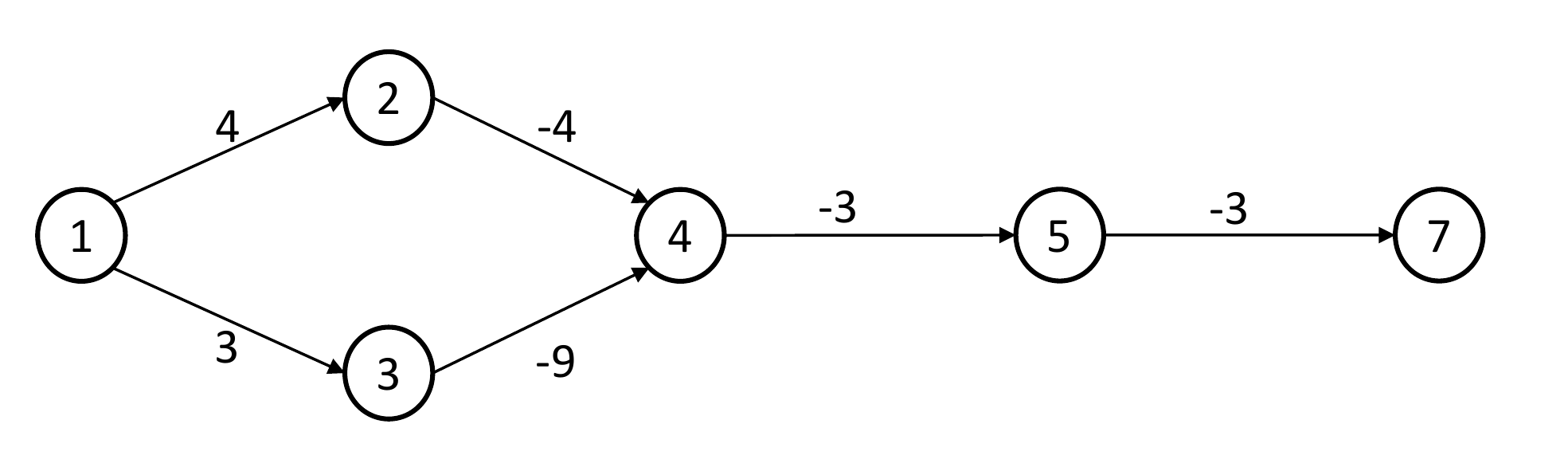}}
	\caption{ (a) HCG of floorplan in Fig. \ref{pic:leg_exmpl}(a), in which vertex $i$ represents module $m_i$, $i=1, 2, \dots, 7$; (b)  horizontal critical relationships of $HCG$ and the weights of the edges.}
	\label{pic:leg_exmpl_HCG}
\end{figure}

Fig. \ref{pic:leg_exmpl_HCG}(a) presents the horizontal constraint graph corresponding to the floorplan in  Fig. \ref{pic:leg_exmpl}(a).
First, according to the current $HCG$ and $VCG$,  line 1 of Algorithm \ref{alg:lg:part} will obtain a floorplan without overlapping that is compressed in the horizontal and vertical directions, which is shown in Fig. \ref{pic:leg_exmpl}(b).

Then because $W_d > W$, we have to legalize the modules in the horizontal direction, such that they are all in the horizontal outline.
First, we need to extract critical relationships from $HCG$ through line 3 of Algorithm \ref{alg:lg:part}, and  line 4 will delete all  compressible relationships until $W_d\leq W$. In this example, since there are no compressible relationships, we use Eq. \eqref{eq:hfh:cal4} to calculate the edge weights. Fig. \ref{pic:leg_exmpl_HCG}(b) is the corresponding critical relationships of $HCG$ and the weights of the edges.

Now we need to select an edge from $HCG$ with the largest weight to delete, and add a new vertical edge in $VCG$. We sequentially delete $m_1 \rightarrow m_2$ and then add $m_1 \uparrow m_2$, delete $m_1 \rightarrow m_3$ and then add $m_1 \uparrow m_3$. It is worth to note that when we delete a relationship and create a new relationship, the weights of related edges will be updated. Finally, after executing line 15 of Algorithm \ref{alg:lg:part}, we get a horizontally legal floorplan, which is shown in Fig. \ref{pic:leg_exmpl}(c). 

\section{Experimental Results}\label{lab:syjg}

In this section, we compare our fixed-outline floorplanning algorithm with
state-of-the-arts experimentally on MCNC \cite{MCNCbench}, GSRC \cite{GSRCbench},
HB+ \cite{2006HBp} and ami49\_x \cite{2008A, 2003Multilevel} benchmarks.
Our algorithm was implemented in C++  and run
on a Linux machine with 3.60GHz Intel Core i3-9100 CPU and 4GB memory.
We focus on floorplanning with hard and soft modules.
In comparisons, wirelength is calculated using the HPWL, and all compared data are cited from relevant literatures directly.

In experiments of our floorplanning algorithm, parameters were set as follows: $k_{max}=1200$
and $O_{min}$ is set to about 1\%
in  Algorithm \ref{alg:n:all}, and $N=10$ in Algorithm \ref{alg:lg:all}.

1) MCNC and GSRC benchmarks: These benchmarks are ami33, ami49, n100, n200 and n300. The numbers of modules are 33, 49, 100, 200 and 300, respectively, and all modules are soft.
We conducted two experiments. The first experiment fixed the whitespace at 15\%,
and changed the aspect ratio of floorplanning area from 1:1 to 4:1.
The purpose is to test the adaptability of the algorithms to aspect ratio.
The second experiment reduced the whitespace to 10\%, and fixed the aspect ratio as 1:1. 
In addition, for the two sets of experiments, we set $\frac{1}{3}=AR^l_i\le \frac{h_i}{w_i}\le AR_i^u=3$ for all soft modules.

In the first experiment, we compare the test results of our floorplanning algorithm using the accelerated partial derivative calculation method (``ours", the same meaning in the sequel) with Analytical \cite{2006AFD}, which is an analytical approach,  and with Parquet-4 \cite{2003Fixed} which is based on simulated annealing. In this experiment, none of the three floorplanners uses multilevel framework. For all test benchmarks, the I/O pads are fixed at the location given originally. The detailed test results are put in Table \ref{tab:1}.

In the table, the first and second columns present the names of the test benchmarks and the floorplanning algorithms, respectively. 	Columns 3-6 give the test results of respective floorplanners. The last three rows of the table show the average results of the two floorplanners proportional to that of our floorplanning algorithm, respectively. From  Table \ref{tab:1}, it can be seen that for different aspect ratio, the average HPWL of our algorithm is at least 16\% less than  that of Analytical \cite{2006AFD},
and at least 31\% less than that of Parquet-4 \cite{2003Fixed}.



\begin{table}[h]
	\centering
	\caption{HPWL results of  three algorithms on the MCNC and GSRC benchmarks, whitespace 15\% }\label{tab:1}
	{\scriptsize
		\begin{tabular}{|c|c|r|r|r|r|}
			\hline
			\multirow{2}{*}{Name}  & \multirow{2}{*}{Algorithm} & \multicolumn{4}{c|}{Aspect ratio $\gamma$}                                                                                     \\ \cline{3-6}
			                       &                            & \multicolumn{1}{c|}{1:1}                   & \multicolumn{1}{c|}{2:1}  & \multicolumn{1}{c|}{3:1}  & \multicolumn{1}{c|}{4:1}  \\ \hline
			\multirow{3}{*}{ami33} & Parquet-4                  & 82149                                      & 79131                     & 91721                     & 101274                    \\ \cline{2-6}
			                       & Analytical                 & 74072                                      & 75168                     & 75180                     & 79529                     \\ \cline{2-6}
			                       & ours                       & 64134                                      & 65094                     & 67457                     & 69615                     \\ \hline
			\multirow{3}{*}{ami49} & Parquet-4                  & 928597                                     & 942117                    & 1092771                   & 1003220                   \\ \cline{2-6}
			                       & Analytical                 & 799239                                     & 829888                    & 880387                    & 939049                    \\ \cline{2-6}
			                       & ours                       & 668608                                     & 674155                    & 727192                    & 761057                    \\ \hline
			\multirow{3}{*}{n100}  & Parquet-4                  & 342103                                     & 351542                    & 351338                    & 392118                    \\ \cline{2-6}
			                       & Analytical                 & 291628                                     & 290158                    & 298894                    & 313060                    \\ \cline{2-6}
			                       & ours                       & 281655                                     & 281294                    & 288535                    & 289199                    \\ \hline
			\multirow{3}{*}{n200}  & Parquet-4                  & 630014                                     & 645219                    & 639803                    & 685057                    \\ \cline{2-6}
			                       & Analytical                 & 572145                                     & 565927                    & 583282                    & 608074                    \\ \cline{2-6}
			                       & ours                       & 509680                                     & 510658                    & 523230                    & 539112                    \\ \hline
			\multirow{3}{*}{n300}  & Parquet-4                  & 770354                                     & 780406                    & 838600                    & 872501                    \\ \cline{2-6}
			                       & Analytical                 & 702822                                     & 722527                    & 793771                    & 858346                    \\ \cline{2-6}
			                       & ours                       & 576158                                     & 575100                    & 602310                    & 656570                    \\ \hline
			\multirow{3}{*}{Ratio} & Parquet-4                  & \multicolumn{1}{c|}{1.31}                  & \multicolumn{1}{c|}{1.33} & \multicolumn{1}{c|}{1.36} & \multicolumn{1}{c|}{1.32} \\ \cline{2-6}
			                       & Analytical                 & \multicolumn{1}{c|}{1.16}                  & \multicolumn{1}{c|}{1.18} & \multicolumn{1}{c|}{1.19} & \multicolumn{1}{c|}{1.21} \\ \cline{2-6}
			                       & ours                       & \multicolumn{1}{c|}{1.00}                  & \multicolumn{1}{c|}{1.00} & \multicolumn{1}{c|}{1.00} & \multicolumn{1}{c|}{1.00} \\ \hline
		\end{tabular}}
\end{table}


In the second experiment, for each benchmark, the aspect ratio was set to $\gamma=1$, and the I/O pads were shifted to the boundary of the floorplanning area. Compared floorplanning algorithms are:
analytical-based methods AR \cite{2008Large} and Ref.\cite{2018LJM} without using multilevel framework, Capo 10.2 \cite{2004capo10.2}, IMF \cite{2008A} and IMF+AMF \cite{2003Multilevel} based on multilevel framework, DeFer \cite{Yan2010DeFer}, and  Parquet-4 \cite{2003Fixed} based on simulated annealing.

Table \ref{tab:1_1} lists the HPWL results obtained by respective algorithms. {\color{blue} The data of DeFer are cited from \cite{Yan2010DeFer},} and the data of the other compared algorithms are cited from Refs. \cite{2008Large} and \cite{2018LJM} directly. On average, our HPWL result {\color{blue} is 6\%,} 6\%, 7\%, 13\% and 29\% less than {\color{blue}that of DeFer \cite{Yan2010DeFer},} IMF \cite{2008A}, IMF+AMF \cite{2003Multilevel}, Capo 10.2 \cite{2004capo10.2} and Parquet-4 \cite{2003Fixed}, respectively.
Compared to the analytical algorithms  Ref.\cite{2018LJM} and AR \cite{2008Large}, our HPWL result is   2\% and 5\% less.
\begin{table}[h]
	\centering
	\caption{HPWLs of  GSRC benchmarks, aspect ratio 1:1, whitespace 10\%}
	\label{tab:1_1}
	\resizebox{\hsize}{!}{
		\footnotesize
		\begin{tabular}{|r|r|r|r|r|r|r|r|r|}
			\hline
			\multicolumn{1}{|c|}{Name}                      &
			\multicolumn{1}{c|}{Parquet-4}                  &
			\multicolumn{1}{c|}{\begin{tabular}[c]{@{}c@{}}Capo\\10.2\end{tabular}} &
			\multicolumn{1}{c|}{AR}                         &
			\multicolumn{1}{c|}{IMF}                        &
			\multicolumn{1}{c|}{\begin{tabular}[c]{@{}c@{}}IMF\\+AFF\end{tabular}} &
			\multicolumn{1}{c|}{Ref. \cite{2018LJM}}        &
			\multicolumn{1}{c|}{\color{blue}DeFer}        &
			\multicolumn{1}{c|}{ours}                                                                                      \\ \hline
			n100                                            & 242050 & 224390 & 203700 & 207852 & 208772 & 198649 & \color{blue}208650  & 194273 \\ \hline
			n200                                            & 432882 & 385594 & 367880 & 369888 & 372845 & 351193 &  \color{blue}372546 & 349301 \\ \hline
			n300                                            & 647452 & 522968 & 492830 & 489868 & 494480 & 483757 &  \color{blue}498909 & 468235 \\ \hline
			Ratio                                           &
			\multicolumn{1}{c|}{1.29}                       &
			\multicolumn{1}{c|}{1.13}                       &
			\multicolumn{1}{c|}{1.05}                       &
			\multicolumn{1}{c|}{1.06}                       &
			\multicolumn{1}{c|}{1.07}                       &
			\multicolumn{1}{c|}{1.02}                       &
			\multicolumn{1}{c|}{\color{blue}1.06}                       &
			\multicolumn{1}{c|}{1.00}                                                                                      \\ \hline
		\end{tabular}
	}
\end{table}

2) HB+ benchmarks:
HB+ benchmarks were generated from HB benchmarks,
with the largest hard macros being enlarged by 100\% and the area of remaining soft macros
reduced to preserve the total cell area.
Therefore, HB+ benchmarks 
are very hard to handle. In this experiment,
we compare the results by our algorithm to multilevel based
floorplanners DeFer \cite{Yan2010DeFer}, QinFer \cite{2021QinFer} and  Ref.\cite{2018LJM}.

DeFer \cite{Yan2010DeFer} has the characteristics of fast speed and high solution quality.
It can solve large-scale fixed-outline slicing floorplanning problems.
QinFer \cite{2021QinFer} is a newly published partitioning based fixed-outline
floorplanning algorithm with high solution quality.
Ref.\cite{2018LJM} is an analytical floorplanner based on multilevel framework. Note that we have two versions of calculating partial derivatives of potential energy. One is the partial derivative calculation method before acceleration (PDBA) presented in Section \ref{lab:pdc}, and another one is
the accelerated partial derivative calculation method presented in Section \ref{lab:sjyh}. Hence we have two versions of floorplanning algorithm respectively. ``ours PDBA" means our floorplanning algorithm using the partial derivative calculation method before acceleration, and ``ours" indicates our floorplanning algorithm using the accelerated partial derivative calculation method.

\begin{table*}[t]
	\caption{Results on the HB+ benchmarks}
	\label{tab:2}
		 \scriptsize
		\centering
		\begin{tabular}{|l|r|r|r|c|r|r|r|r|r|r|r|r|r|r|}
			\hline
			\multicolumn{5}{|c|}{Basic Information}                        &
			\multicolumn{5}{c|}{HPWL ($10^6$)}                             &
			\multicolumn{5}{c|}{time (s)}                                                                                                                                             \\ \hline
			\multicolumn{1}{|c|}{\rotatebox{0}{Name} }                     &
			\multicolumn{1}{c|}{\rotatebox{0}{Hard }}                      &
			\multicolumn{1}{c|}{\rotatebox{0}{Soft }}                      &
			\multicolumn{1}{c|}{I/O}                                       &
			\multicolumn{1}{c|}{\rotatebox{0}{\begin{tabular}[c]{@{}c@{}} white-\\ space\end{tabular}}} &
			\multicolumn{1}{c|}{\rotatebox{0}{QinFer}}                     &
			\multicolumn{1}{c|}{\rotatebox{0}{Ref. \cite{2018LJM}}}        &
			\multicolumn{1}{c|}{\rotatebox{0}{DeFer}}                      &
			\multicolumn{1}{c|}{\rotatebox{0}{\begin{tabular}[c]{@{}c@{}}ours\\ PDBA\end{tabular}}} &
			\multicolumn{1}{c|}{\rotatebox{0}{\begin{tabular}[c]{@{}c@{}}ours\end{tabular}}} &
			\multicolumn{1}{c|}{\rotatebox{0}{QinFer}}                     &
			\multicolumn{1}{c|}{\rotatebox{0}{Ref. \cite{2018LJM}}}        &
			\multicolumn{1}{c|}{\rotatebox{0}{DeFer}}                      &
			\multicolumn{1}{c|}{\rotatebox{0}{\begin{tabular}[c]{@{}c@{}} ours\\ PDBA\end{tabular}}} &
			\multicolumn{1}{c|}{\rotatebox{0}{\begin{tabular}[c]{@{}c@{}} ours\end{tabular}}}                                                                                                            \\ \hline
			HB+01                                                          & 665  & 246 & 246 & 26\% & 2.91   & 3.18   & 3.09   & 3.03   & 3.00   & 5.0  & 150 & 1.8  & 19.49  & 4.3  \\ \hline
			HB+02                                                          & 1200 & 271 & 259 & 25\% & 6.47   & 6.80   & 6.17   & 6.53   & 6.02   & 11.3 & 338 & 15.3 & 100.05 & 4.3  \\ \hline
			HB+03                                                          & 999  & 290 & 283 & 30\% & 9.17   & 9.68   & 9.19   & 8.27   & 8.17   & 8.6  & 262 & 4.0  & 88.77  & 8.1  \\ \hline
			HB+04                                                          & 1289 & 295 & 287 & 25\% & 11.16  & 9.87   & 10.26  & 10.03  & 9.72   & 12.4 & 209 & 14.2 & 137.16 & 7.5  \\ \hline
			HB+06                                                          & 571  & 178 & 166 & 25\% & 8.03   & 8.50   & 8.78   & 7.95   & 7.91   & 7.8  & 125 & 5.0  & 24.36  & 7.1  \\ \hline
			HB+07                                                          & 829  & 291 & 287 & 25\% & 14.88  & 15.10  & 15.48  & 14.73  & 14.07  & 8.5  & 230 & 4.6  & 74.21  & 7.2  \\ \hline
			HB+08                                                          & 968  & 301 & 286 & 26\% & 16.49  & 17.60  & 18.73  & 18.35  & 17.19  & 22.8 & 324 & 19.3 & 73.76  & 9.7  \\ \hline
			HB+09                                                          & 860  & 253 & 285 & 25\% & 17.24  & 18.30  & 16.66  & 16.32  & 15.81  & 8.7  & 259 & 4.2  & 58.82  & 9.0  \\ \hline
			HB+10                                                          & 809  & 786 & 744 & 20\% & 42.33  & 46.70  & 45.12  & 41.58  & 40.61  & 11.4 & 191 & 6.3  & 151.73 & 32.9 \\ \hline
			HB+11                                                          & 1124 & 373 & 406 & 25\% & 25.7   & 28.20  & 26.99  & 25.38  & 24.58  & 10.1 & 494 & 7.1  & 77.92  & 9.3  \\ \hline
			HB+12                                                          & 582  & 651 & 637 & 26\% & 52.83  & 53.60  & 50.17  & 49.26  & 48.96  & 16.6 & 136 & 5.5  & 54.36  & 15.0 \\ \hline
			HB+13                                                          & 830  & 424 & 490 & 25\% & 35.44  & 35.40  & 35.51  & 32.51  & 32.65  & 23.7 & 90  & 5.9  & 41.56  & 14.2 \\ \hline
			HB+14                                                          & 1021 & 614 & 517 & 25\% & 60.68  & 63.40  & 64.50  & 61.57  & 59.62  & 14.9 & 208 & 12.0 & 76.22  & 16.7 \\ \hline
			HB+15                                                          & 1019 & 393 & 383 & 25\% & 77.48  & 79.10  & 84.29  & 74.97  & 73.93  & 37.1 & 532 & 14.7 & 89.37  & 24.8 \\ \hline
			HB+16                                                          & 633  & 458 & 504 & 25\% & 98.53  & 92.90  & 98.66  & 87.66  & 87.32  & 11.2 & 160 & 8.1  & 47.49  & 20.0 \\ \hline
			HB+17                                                          & 682  & 760 & 743 & 25\% & 140.84 & 140.00 & 144.56 & 136.37 & 138.44 & 15.2 & 169 & 14.7 & 51.71  & 20.5 \\ \hline
			HB+18                                                          & 658  & 285 & 272 & 25\% & 70.36  & 70.70  & 71.86  & 68.76  & 68.05  & 12.4 & 66  & 11.3 & 32.84  & 14.5 \\ \hline
			\multicolumn{5}{|c|}{Ratio}                                    &
			\multicolumn{1}{c|}{1.05}                                      &
			\multicolumn{1}{c|}{1.08}                                      &
			\multicolumn{1}{c|}{1.08}                                      &
			\multicolumn{1}{c|}{1.02}                                      &
			\multicolumn{1}{c|}{1.00}                                      &
			\multicolumn{1}{c|}{1.23}                                      &
			\multicolumn{1}{c|}{24.43}                                     &
			\multicolumn{1}{c|}{0.89}                                      &
			\multicolumn{1}{c|}{7.06}                                      &
			\multicolumn{1}{c|}{1.00}                                                                                                                                                 \\ \hline
		\end{tabular}
\end{table*}

Table \ref{tab:2} gives the basic information of the HB+ benchmarks
and the test results of  compared floorplanners.
The first 5 columns of Table \ref{tab:2} present the basic information of the benchmarks,
including the names of benchmarks, the numbers of hard modules and soft modules,
the number of terminals (I/O pads) and the whitespace fractions.
The other information of each benchmark, such as the width and height of floorplanning area,
the upper and lower bounds of aspect ratios of soft modules,
can be found in the corresponding literature. In the table, columns 6-10 are the HPWL results obtained by the respective algorithms.
According to the last row of Table \ref{tab:2},
the average HPWL of our algorithm is 5\%, 8\% and 8\% less than that of
QinFer, DeFer and Ref.\cite{2018LJM},  respectively.

In Table \ref{tab:2}, columns 11-15 present the run-times of different algorithms, in which the run-times of DeFer,  Ref.\cite{2018LJM} and QinFer are directly cited from \cite{Yan2010DeFer}, \cite{2018LJM} and \cite{2021QinFer}, respectively. The computing environment of DeFer was Core Duo 1.86GHz CPU with 2GB memory,
QinFer was Intel core i5-7500 3.40GHz with 8GB memory,
and Ref.\cite{2018LJM} was Intel Xeon 2.00GHz CPU.
On average, the run-times of DeFer and QinFer are 89\% and 123\% of our algorithm,
respectively. However, the run-time of Ref. \cite{2018LJM} is quite long.
Although the computing environments are different,
the speed of our floorplanning algorithm can still be compared roughly.

Note that both of our floorplanning algorithm and Ref. \cite{2018LJM} are modified from VLSI placement algorithms respectively, and both of them do not use multi-level framework for the small-scale benchmarks in Table \ref{tab:1_1}. According to Table \ref{tab:1_1}, the average HPWL of our algorithm  is  2\% less than that of Ref.\cite{2018LJM}.
However, for  large-scale benchmarks in Table \ref{tab:2},	our floorplanning algorithm still does not use  multi-level framework, but the algorithm in Ref.\cite{2018LJM} does.  According to Table \ref{tab:2},
the HPWL of our algorithm is  8\% less than that of Ref. \cite{2018LJM}. 	
This further improvement is obviously due to that our floorplanning algorithm is flat, i.e.,  does not use the multi-level framework.


Theoretically, the results of ``our PDBA" should be better than ``ours", since it calculates exactly the partial derivatives of the potential energy. However, according to Table \ref{tab:2}, the average HPWL result of ``our PDBA" has a slight increase of 2\% over ``ours". This seems to be inconsistent with the theory. In fact, the fixed-outline floorplanning is a non-convex optimization problem, and its objective function has multiple local minima. Therefore, a better solution may not be obtained by searching in the exact gradient descent direction. While, each iteration in ``ours" has a certain error of partial derivative calculation, which makes our algorithm have limited ability to jump out of local optima, and find a better solution.


3) ami49\_x benchmarks:
In this experiment, we test our  floorplanning algorithm on ami49\_x benchmarks \cite{2008A, 2003Multilevel} to see its performance on large-scale fixed-outline floorplanning.
The benchmarks ami49\_x were designed as follows:
First, duplicate each module $v_i$ and net $x$ times, and get new modules $v_{i,1}, v_{i,2}, \dots, v_{i,x}$. Then, add $x-1$ nets $(v_{i,1}, v_{i,2}), (v_{i,1}, v_{i,3})$, $\dots$, $(v_{i,1}$, $v_{i,x})$.
Finally, reduce the width and height of each module by 5 times
to avoid overflow in computing the wirelength. In this experiment, the whitespace is  15\%, the aspect ratio of floorplanning area is  $\gamma=1$, and the I/O pads are shifted to the boundary of the floorplanning area.

Table \ref{tab:3} presents the test results of our algorithm,
Capo 10.2 \cite{2004capo10.2}, MB*-tree \cite{2008A}, and IMF+AFF \cite{2003Multilevel}. The first three columns of Table \ref{tab:3} give the basic information.
Columns 4-7 and 8-11 are the HPWL and run-time results of each algorithm.
The computing environments of Capo 10.2, MB*-tree and IMF+AFF were
Intel Pentium-4 3.20GHz CPU with 3GB memory. The test results of Capo 10.2, MB*-tree and IMF+AFF are cited from \cite{2003Multilevel} directly. Note that, our algorithm does not employ multilevel framework used in the compared floorplanners.

\begin{table}[h]\centering
	\caption{Results on the ami49\_x benchmarks, whitespace 15\%}
	\label{tab:3}
	\resizebox{\hsize}{!}{
		\begin{tabular}{|l|r|r|r|r|r|r|r|r|r|r|}
			\hline
			\multicolumn{3}{|c|}{Basic Information}         & \multicolumn{4}{c|}{wirelength ($10^6$)}
			                                                & \multicolumn{4}{c|}{time (min)}                                                                                   \\ \hline
			\multicolumn{1}{|c|}{Name}                      &
			\multicolumn{1}{c|}{ Module}                    &
			\multicolumn{1}{c|}{Net}                        &
			\multicolumn{1}{c|}{\begin{tabular}[c]{@{}c@{}}MB*-\\ tree\end{tabular}} &
			\multicolumn{1}{c|}{\begin{tabular}[c]{@{}c@{}}IMF\\ +AFF\end{tabular}} &
			\multicolumn{1}{c|}{\begin{tabular}[c]{@{}c@{}}Capo\\ 10.2\end{tabular}} &
			\multicolumn{1}{c|}{ours}                       &
			\multicolumn{1}{c|}{\begin{tabular}[c]{@{}c@{}}MB*-\\ tree\end{tabular}} &
			\multicolumn{1}{c|}{\begin{tabular}[c]{@{}c@{}}IMF\\ +AFF\end{tabular}} &
			\multicolumn{1}{c|}{\begin{tabular}[c]{@{}c@{}}Capo\\ 10.2\end{tabular}} &
			\multicolumn{1}{c|}{ours}                                                                                                                                           \\ \hline
			ami49\_10                                       & 490                                      & 4521  & 3.79   & 3.31   & 3.20   & 2.68   & 2.1   & 0.2  & 0.9  & 0.08 \\ \hline
			ami49\_20                                       & 980                                      & 9091  & 10.20  & 8.13   & 7.66   & 6.80   & 4.8   & 0.5  & 1.9  & 0.09 \\ \hline
			ami49\_40                                       & 1960                                     & 18231 & 26.70  & 20.20  & 19.30  & 19.03  & 12.7  & 1.3  & 3.7  & 0.22 \\ \hline
			ami49\_60                                       & 2940                                     & 27371 & 49.90  & 34.20  & 33.70  & 32.30  & 25.7  & 2.2  & 6.4  & 0.43 \\ \hline
			ami49\_80                                       & 3920                                     & 36511 & 74.80  & 50.50  & 49.50  & 47.30  & 36.7  & 3.5  & 7.8  & 1.08 \\ \hline
			ami49\_100                                      & 4900                                     & 45651 & 103.00 & 68.60  & 67.90  & 65.32  & 66.2  & 5.1  & 10.4 & 1.57 \\ \hline
			ami49\_150                                      & 7350                                     & 68501 & 184.00 & 121.00 & 119.00 & 113.86 & 107.5 & 11.0 & 18.4 & 2.57 \\ \hline
			ami49\_200                                      & 9800                                     & 91351 & 334.00 & 180.00 & 177.00 & 170.02 & 224.8 & 19.1 & 22.4 & 5.02 \\ \hline
			\multicolumn{3}{|c|}{Ratio}                     &
			\multicolumn{1}{c|}{1.58}                       &
			\multicolumn{1}{c|}{1.10}                       &
			\multicolumn{1}{c|}{1.07}                       &
			\multicolumn{1}{c|}{1.00}                       &
			\multicolumn{1}{c|}{43.43}                      &
			\multicolumn{1}{c|}{3.88}                       &
			\multicolumn{1}{c|}{6.50}                       &
			\multicolumn{1}{c|}{1.00}                                                                                                                                           \\ \hline
		\end{tabular}
	}
\end{table}

From the last row of Table \ref{tab:3}, the average HPWL of our  floorplanning algorithm is 7\%, 10\% and 58\% less than that of Capo 10.2, IMF+AFF and MB*-tree, respectively.
In terms of run-time, we cannot make direct comparisons due to different computing environment.
Despite that, our floorplanning algorithm can succeed in solving every instance with at most 9800 modules in about no more than 5  minutes.
While, the compared floorplanners require at least 19 minutes of computing. Therefore, the run-time of our algorithm is acceptable.

\begin{figure}[h]
	\centering
	\includegraphics[width=0.75\hsize]{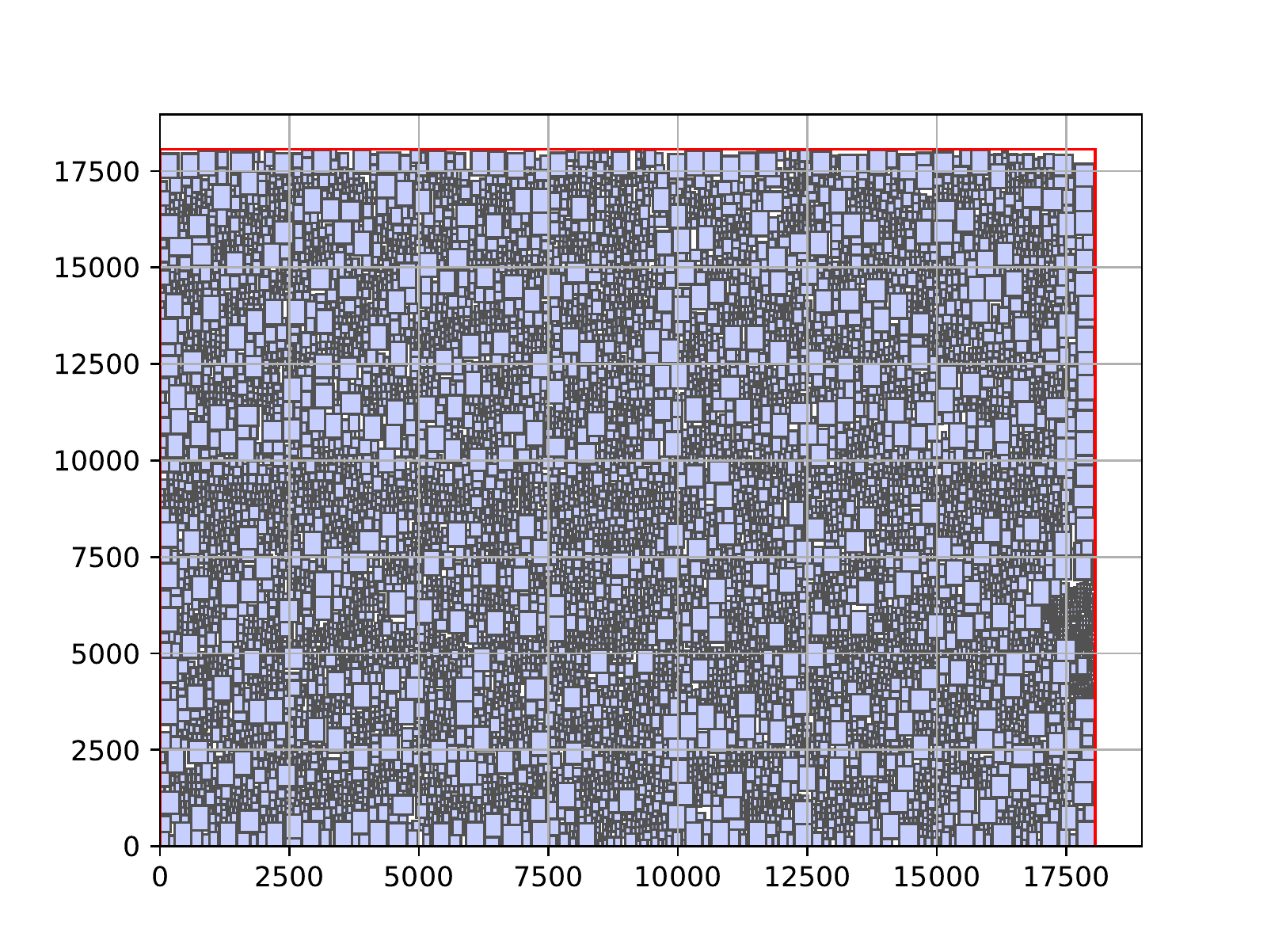}
	\caption{Result of ami49\_200 by our floorplanning algorithm (9800 modules, 91351 nets, $HPWL=170.02\times 10^6$).}
	\label{pic:ami49-200}
\end{figure}

Fig. \ref{pic:ami49-200} shows the results of our algorithm on the largest test case ami49\_200. It can be seen that the quality of our algorithm is reliable.

4) Run-time analysis:
Next, we make the running time component analysis of our floorplanning algorithm with accelerated partial derivative calculation method.
Note that, the total running time of our algorithm $=$ initial solution generation time by QP $+$  run-time of our global floorplanning $+$ run-time of our legalization.

\begin{figure}[h]
	\centering
	\includegraphics[width=0.9\hsize]{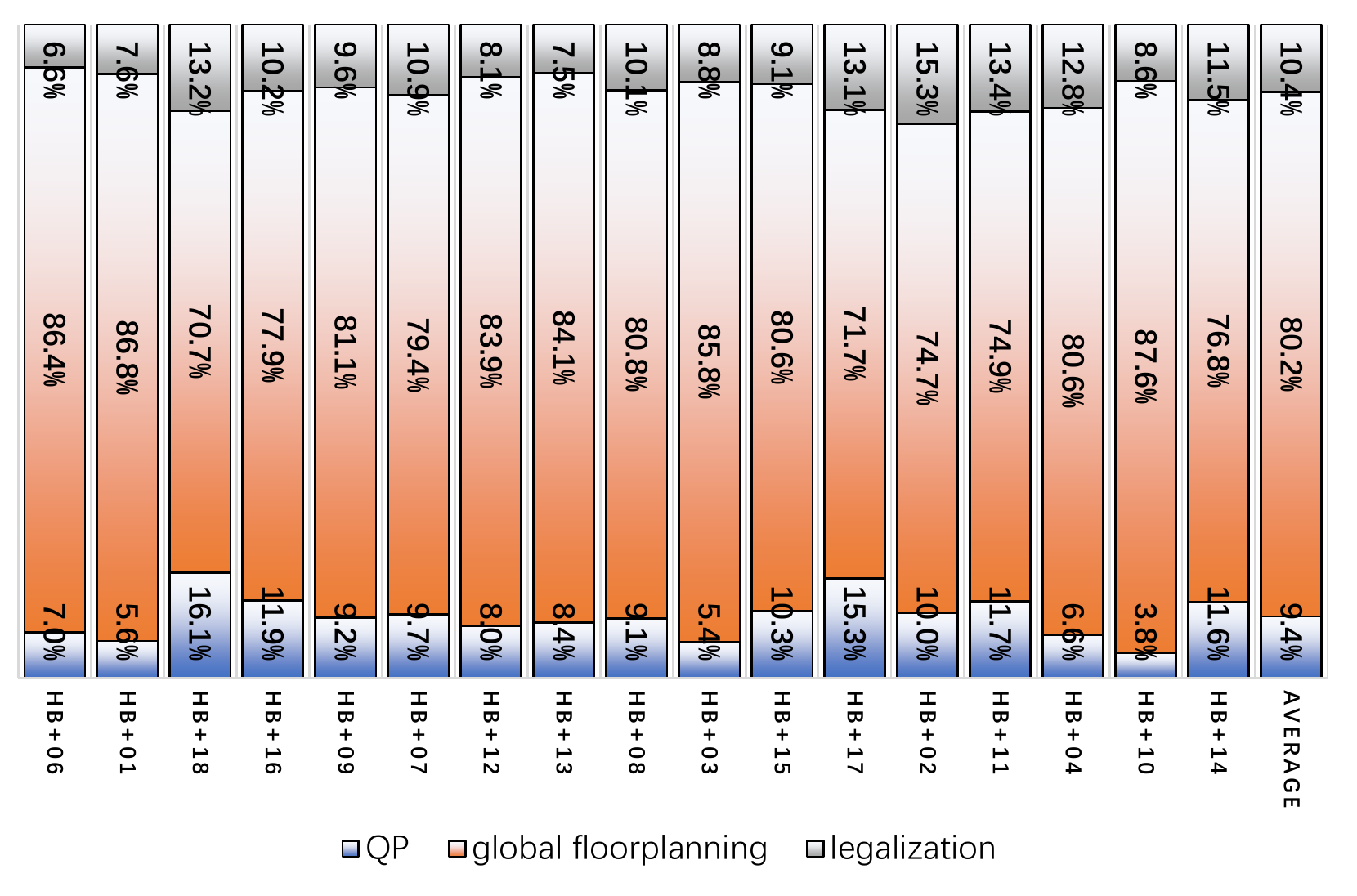}
	\caption{Proportions of run-time spent by QP, global floorplanning and legalization on the HB+ benchmarks.}
	\label{pic:timep}
\end{figure}

Fig. \ref{pic:timep} presents the run-time proportion of our algorithm on all  HB+ benchmarks. In the figure, the abscissa represents the benchmark name,
sorted from the left to right in increasing scale. For each benchmark, from the bottom to top, the figure shows the run-time proportions of QP, global floorplanning and legalization, respectively. On average, our floorplanning algorithm uses 9.4\% of run-time on generating initial solution, uses 80.2\% of run-time on global floorplanning, and uses 10.4\% of run-time on legalization.	Therefore, our global floorplanning consumes most of run-time among the three components.

\begin{figure}[h]
	\centering
	\includegraphics[width=0.9\hsize]{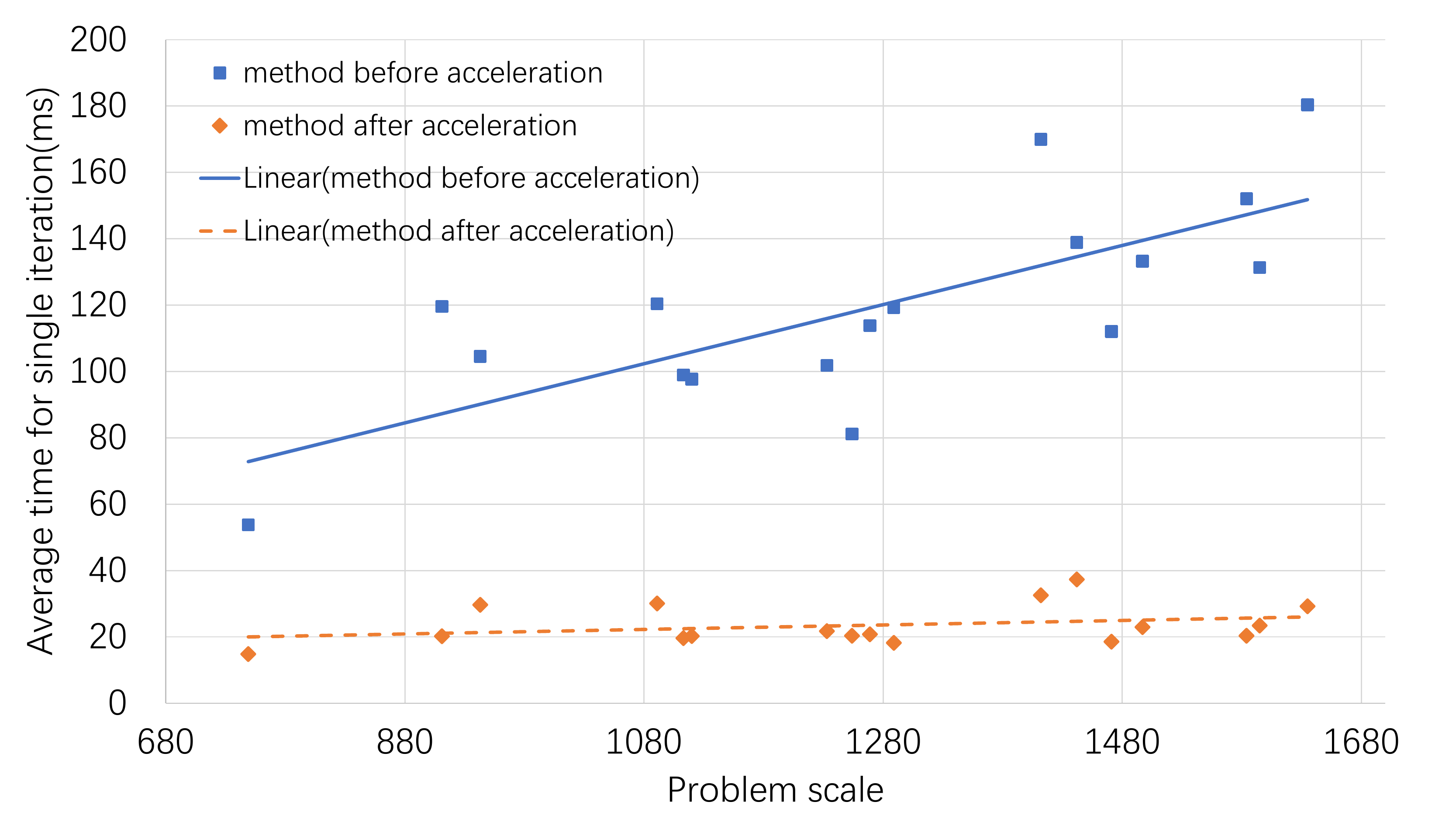}
	\caption{Relationship between the average run-time of a single iteration of our global floorplanning algorithm with exact and with approximate partial derivative calculation methods.}
	\label{pic:time}
\end{figure}

Since the run-time of our global floorplanning is approximately equal to the number of global floorplanning iterations $\times$ the average run-time of a single iteration of the global floorplanning. Fig. \ref{pic:time} gives the trend  chart of the average run-time of a single iteration of our global floorplanning on HB+ benchmarks with respect to the benchmark scale.

In the figure, the solid line and the dashed line are the liner fittings of the average run-times of a single iteration of our global floorplanning
algorithm using exact partial derivative calculation method, and the approximate partial derivative calculation method, respectively. It can be seen that both of them have linear time complexity, but the latter one is much faster.


5) Results of global floorplanning and legalization on HB+  and ami49\_x benchmarks: Next, we analyze the overlap rate after global floorplanning and the wire-length increase after legalization of our floorplanning algorithm.
Table \ref{tab:4} gives the overlap rates (columns 2, 7) and HPWLs (columns 3, 8) before legalization, the  HPWLs after legalization (columns 4, 9),  	and the wirelength increase of HPWL (columns 5, 10) after our legalization algorithm for each benchmark, respectively.
	 Here, the overlap rate is calculated by
		$$
			Overlap\ Rate=\sum_{i=0}^{K-1}\sum_{j=0}^{K-1}\max \{ OR(i,j),\  0 \} \times 100\%,
		$$
in which
		$$
			OR(i,j)=\frac{\sum_{v_k\in V}Area(bin_{i,j}\cap R_k)}{Area(bin_{i,j})}-1.
		$$
The wirelength increase of HPWL is calculated using
$$\frac {HPWL~ After~ Legal. - HPWL~ Before~ Legal. }{ HPWL~Before~ Legal.}\times 100\%,$$
for which a positive number means that the legalization has increased wirelength, and a negative number means that wirelength has been decreased.

\begin{table}[h]
	\caption{Results of overlap rate after global floorplanning and HPWL increase after legalization}
	\label{tab:4}
	\resizebox{\hsize}{!}{
		\begin{tabular}{|c|r|rr|r|c|r|rr|r|}
			\hline
			\multirow{2}{*}{Name}                                              &
			\multicolumn{1}{c|}{\multirow{2}{*}{\begin{tabular}[c]{@{}c@{}}Overlap\\ rate\end{tabular}}}   &
			\multicolumn{2}{c|}{ HPWL($10^6$)}                                 &
			\multicolumn{1}{c|}{\multirow{2}{*}{ \begin{tabular}[c]{@{}c@{}}Rate of\\ change\end{tabular} }} &
			\multirow{2}{*}{Name}                                              &
			\multicolumn{1}{c|}{\multirow{2}{*}{\begin{tabular}[c]{@{}c@{}}Overlap\\ rate\end{tabular}}}   &
			\multicolumn{2}{c|}{HPWL($10^6$)}                                  &
			\multicolumn{1}{c|}{\multirow{2}{*}{\begin{tabular}[c]{@{}c@{}}Rate of\\ change\end{tabular}}}                                                                                                                                                                                           \\ \cline{3-4} \cline{8-9}
			                                                                   &
			\multicolumn{1}{c|}{}                                              &
			\multicolumn{1}{c|}{Before }                                       &
			\multicolumn{1}{c|}{After }                                        &
			\multicolumn{1}{c|}{}                                              &
			                                                                   &
			\multicolumn{1}{c|}{}                                              &
			\multicolumn{1}{c|}{Before}                                        &
			\multicolumn{1}{c|}{After}                                         &
			\multicolumn{1}{c|}{}                                                                                                                                                                                                                                      \\ \hline
			HB+01                                                              & 0.97\% & \multicolumn{1}{r|}{2.99}  & 3.00  & 0.48\%  & HB+15      & 1.00\% & \multicolumn{1}{r|}{75.18}                               & 73.93                              & -1.66\% \\ \hline
			HB+02                                                              & 1.00\% & \multicolumn{1}{r|}{6.05}  & 6.02  & -0.42\% & HB+16      & 1.07\% & \multicolumn{1}{r|}{88.54}                               & 87.32                              & -1.38\% \\ \hline
			HB+03                                                              & 0.94\% & \multicolumn{1}{r|}{8.23}  & 8.17  & -0.63\% & HB+17      & 1.02\% & \multicolumn{1}{r|}{141.38}                              & 138.44                             & -2.07\% \\ \hline
			HB+04                                                              & 0.93\% & \multicolumn{1}{r|}{9.80}  & 9.72  & -0.86\% & HB+18      & 1.06\% & \multicolumn{1}{r|}{69.47}                               & 68.05                              & -2.04\% \\ \hline
			HB+06                                                              & 0.92\% & \multicolumn{1}{r|}{8.02}  & 7.91  & -1.33\% & ami49\_10  & 0.95\% & \multicolumn{1}{r|}{2.69}                                & 2.68                               & -0.36\% \\ \hline
			HB+07                                                              & 0.99\% & \multicolumn{1}{r|}{14.09} & 14.07 & -0.12\% & ami49\_20  & 0.98\% & \multicolumn{1}{r|}{6.71}                                & 6.80                               & 1.20\%  \\ \hline
			HB+08                                                              & 0.97\% & \multicolumn{1}{r|}{17.44} & 17.16 & -1.64\% & ami49\_40  & 0.99\% & \multicolumn{1}{r|}{18.88}                               & 19.03                              & 0.77\%  \\ \hline
			HB+09                                                              & 0.93\% & \multicolumn{1}{r|}{16.06} & 15.81 & -1.50\% & ami49\_60  & 1.00\% & \multicolumn{1}{r|}{31.98}                               & 32.30                              & 1.00\%  \\ \hline
			HB+10                                                              & 1.08\% & \multicolumn{1}{r|}{40.57} & 40.61 & 0.10\%  & ami49\_80  & 1.00\% & \multicolumn{1}{r|}{46.84}                               & 47.30                              & 0.99\%  \\ \hline
			HB+11                                                              & 1.07\% & \multicolumn{1}{r|}{24.58} & 24.58 & -0.03\% & ami49\_100 & 0.70\% & \multicolumn{1}{r|}{65.03}                               & 65.32                              & 0.44\%  \\ \hline
			HB+12                                                              & 0.93\% & \multicolumn{1}{r|}{48.30} & 48.96 & 1.37\%  & ami49\_150 & 0.70\% & \multicolumn{1}{r|}{113.58}                              & 113.86                             & 0.25\%  \\ \hline
			HB+13                                                              & 0.95\% & \multicolumn{1}{r|}{32.93} & 32.65 & -0.85\% & ami49\_200 & 0.89\% & \multicolumn{1}{r|}{168.79}                              & 170.02                             & 0.73\%  \\ \hline
			HB+14                                                              & 1.09\% & \multicolumn{1}{r|}{60.21} & 59.62 & -0.99\% & Average    & 0.96\% & \multicolumn{1}{r|}{\diagbox[dir=NW,height=1em]{~}{~~} } & \diagbox[dir=NW,height=1em]{~}{~~} & -0.34\% \\ \hline
		\end{tabular}
	}
\end{table}

From Table \ref{tab:4}, it can be seen that our global floorplanning optimizes the wirelength while effectively reduces the overlap. On average, the average overlap rate is 0.96\% after global floorplanning.
Further,
our legalization reduces on average the wirelength by 0.34\%. The reduction in wirelength in fact is due to  compressing the floorplan. Note that,
there is no significant change in wirelength before and after legalization. This is due to that our legalization is designed as a local modification algorithm which does not modify massively the global floorplanning result.

\section{Conclusion}\label{conclu}

Based on  a novel mathematical model for characterizing non-overlapping of modules and an analytical solution of Poisson's equation, this paper has proposed an algorithm without multilevel framework for fixed-outline floorplanning. The algorithm consists of global floorplanning  and legalization phases.
In global floorplanning, it uses redefined potential energy based on the novel mathematical model and the analytical solution of Poisson's equation, to realize  spreading of modules and optimizing the widths of soft modules.
In legalization, the algorithm uses constraint graphs to alternatively
eliminate overlaps between modules in the horizontal and vertical directions. Experimental results on the MCNC, GSRC, HB+ and ami49\_x benchmarks show that the average HPWLs by our algorithm
are  at least 2\% and 5\% less than compared state-of-the-arts
on small and large-scale benchmarks respectively,
within  acceptable run-time. Considering other issues in our floorplanning algorithm, such as routability, timing and thermal, are under investigation.

\bibliographystyle{IEEEtran}
\bibliography{IEEEabrv,ref}

\newpage

%
%
%
%
%
%
%
%

\appendix
\section{Appendix}\label{appndx}
\setcounter{lemma}{0}
\setcounter{theorem}{0}
\begin{lemma}
	Any two different modules $v_i$ and $v_j\in V$ are non-overlapping
	if and only if $\iint_{R_i}\rho_j(u,v)\mathrm{d}u \mathrm{d}v=0$.
\end{lemma}
\begin{proof}
	Since $\rho_j(u,v) \ge 0$,  $\iint_{R_i}\rho_j(u,v)\mathrm{d}u \mathrm{d}v=0$ is equivalent to
	$\rho_j(u,v)=0$, for all $(u,v)\in R_i$. This is further equivalent to
	$R_i\cap R_j=\emptyset$, i.e., modules $v_i$ and $v_j\in V$ are non-overlapping.
\end{proof}

\begin{lemma}
	$\iint_{R_i}\rho_i(u,v)\mathrm{d}u \mathrm{d}v=Area(R_i)$  for all modules $v_i\in V$.
\end{lemma}
\begin{proof}
	According to the definition of $\rho_i(u,v)$, it holds that
	$$\iint_{R_i}\rho_i(u,v)\mathrm{d}u \mathrm{d}v=\iint_{R_i}1\mathrm{d}u \mathrm{d}v=Area(R_i).$$
\end{proof}


Based on Lemmas \ref{lem:1} and \ref{lem:2},
the following theorem can be proved.
\begin{theorem}
	Any module $v_i\in V$ is non-overlapping with other modules if and only if
	\begin{equation}\label{equii}
		\iint_{R_i}\sum_{v_j\in V}\rho_j(u,v)\mathrm{d}u \mathrm{d}v=Area(R_i).
	\end{equation}
\end{theorem}
\begin{proof}
	If module $v_i\in V$ is non-overlapping with other modules,
	then by Lemmas \ref{lem:1} and \ref{lem:2},
	\begin{equation*}
		\begin{aligned}
			& \iint_{R_i}\sum_{v_j\in V}\rho_j(u,v)\mathrm{d}u \mathrm{d}v
			=\sum_{v_j\in V}\iint_{R_i}\rho_j(u,v)\mathrm{d}u \mathrm{d}v    \\
			= & \iint_{R_i}\rho_i(u,v)\mathrm{d}u \mathrm{d}v+\sum_{
				v_j\in V \atop
				v_j\neq v_i
			}\iint_{R_i}\rho_j(u,v)\mathrm{d}u \mathrm{d}v                   \\
			= & Area(R_i)+0=Area(R_i).
		\end{aligned}
	\end{equation*}
	
	Conversely, if Eq. \eqref{equii} holds, we can have
	\begin{equation*}\label{thm:1:suf}
		\begin{aligned}
			& Area(R_i)
			=\iint_{R_i}\sum_{v_j\in V}\rho_j(u,v)\mathrm{d}u \mathrm{d}v                                  \\
			= & \iint_{R_i}\rho_i(u,v)\mathrm{d}u \mathrm{d}v
			+\sum_{
				v_j\in V\atop v_j\neq v_i
			}\iint_{R_i}\rho_j(u,v)\mathrm{d}u \mathrm{d}v                                                 \\
			= & Area(R_i)+\sum_{ v_j\in V\atop v_j\neq v_i }\iint_{R_i}\rho_j(u,v)\mathrm{d}u \mathrm{d}v,
		\end{aligned}
	\end{equation*}
	where the last equality comes from  Lemma \ref{lem:2}. Then
	$$\sum_{ v_j\in V\atop v_j\neq v_i }\iint_{R_i}\rho_j(u,v)\mathrm{d}u \mathrm{d}v=0.$$
	Since $\rho_j(u,v)\ge 0$,
	the above equality implies that $\iint_{R_i}\rho_j(u,v)\mathrm{d}u \mathrm{d}v=0$
	holds for modules $v_j\neq v_i$. Hence by  Lemma \ref{lem:1},
	the conclusion holds.
\end{proof}

\end{document}